\documentclass[aps,prb,twocolumn,showpacs,citeautoscript,amsmath,amssymb,floatfix]{revtex4-2}
\usepackage{amsmath,amssymb,color}
\usepackage{bm}
\usepackage{graphicx}
\usepackage{psfrag}
\usepackage{hyperref}
\newcommand{\bd}{\bm}

\begin{document}

\title{Spin functional renormalization group for the $J_1 J_2 J_3$ 
quantum Heisenberg model}

\author{Dmytro Tarasevych, Andreas R\"{u}ckriegel,  Savio Keupert, Vasilios Mitsiioannou, and Peter Kopietz}
  
\affiliation{Institut f\"{u}r Theoretische Physik, Universit\"{a}t
  Frankfurt,  Max-von-Laue Stra{\ss}e 1, 60438 Frankfurt, Germany}

\date{February 27, 2023}

\begin{abstract}
We use our recently developed functional renormalization
group (FRG) approach for quantum spin systems
to investigate the phase diagram
of the frustrated 
$J_1 J_2 J_3$ quantum Heisenberg model on a cubic lattice.
From a simple truncation of the hierarchy of FRG
flow equations for the irreducible spin-vertices
which retains only static spin fluctuations
and neglects the flow of the four-spin interaction,
we can estimate  
the critical temperature with a similar accuracy as
the numerically more expensive pseudofermion FRG. 
In the regime where the ground state exhibits either ferromagnetic or antiferromagnetic order, 
a more  sophisticated truncation including the renormalization 
of the four-spin interaction as well as dynamic spin fluctuations
reveals the underlying renormalization group fixed point
and yields critical temperatures which 
deviate from the accepted values by at most $4 \%$.
\end{abstract}


\maketitle

\section{Introduction}

Frustrated spin systems continue to be of great interest in modern condensed matter physics
because they can exhibit intriguing complexity which can be explored experimentally
and poses a challenge to theory \cite{Diep20}.
While in reduced dimensions spin systems 
can be investigated
with the help of  powerful numerical
methods such as the density matrix renormalization group \cite{White92,Schollwoeck05},  
controlled methods for realistic three dimensional systems are not available.
It is therefore important to develop approximate methods for studying 
frustrated spin systems in an
unbiased way.  
A promising method is the functional renormalization group (FRG)
\cite{Wetterich93,Berges02,Pawlowski07,Kopietz10,Metzner12,Dupuis21}, which
is one of the few unbiased quantitative methods for 
investigating  the dominant instabilities of  interacting 
fermions on a lattice \cite{Metzner12,Dupuis21,Hille20}.
Given the fact that with the help of Abrikosov's pseudofermion representation of
spin $1/2$ operators \cite{Abrikosov65} 
any spin $1/2$ Hamiltonian can be mapped onto
an interacting fermion Hamiltonian acting on a projected Hilbert space, it is clear that
the established FRG machinery for interacting fermions can also be applied to
quantum spin systems. This strategy, called
pseudofermion FRG,  was 
pioneered by Reuther and W\"{o}lfle \cite{Reuther10} in 2010; since then it has has been used 
to investigate the phase diagrams
and static correlation functions of many different spin models~\cite{Reuther11,Reuther11a,Buessen16,Iqbal16,Baez2017,Rueck18,Thoenniss20,Kiese22,Ritter22}.
However, the pseudofermion FRG has some disadvantages: 
(1) the Hilbert space projection 
can only be implemented approximately, (2) the calculation of the
spin dynamics has so far not been possible,
(3) available truncations are not sufficient to obtain proper renormalization group fixed points and the associated critical behavior, and (4)
explicit solutions of even severely truncated  flow equations 
require heavy numerical calculations. 
At this point,
we should mention that
an alternative representation of the spin $1/2$ operators
in terms of Majorana fermions \cite{Martin1959,Tsvelik1992} has recently been used to
develop a pseudo-Majorana FRG \cite{Niggemann2021,Niggemann22},
which does not suffer from the problem (1) and can also reproduce the non-trivial scaling 
characteristic for a proper renormalization group fixed point \cite{Niggemann22}.
However, the other problems mentioned above remain also for the pseudo-Majorana FRG. 
Motivated by the desire to avoid at least some of these problems,
in Ref.~[\onlinecite{Krieg19}]
an alternative FRG approach to quantum spin systems 
has been proposed which does not rely on any representation of
spin operators in terms of fermionic or bosonic auxiliary operators.
The crucial insight of  Ref.~[\onlinecite{Krieg19}]
is that the  generating functional ${\cal{G}} [ \bd{h} ]$ 
of the imaginary time ordered spin correlation functions
satisfies an exact flow equation which
can be directly obtained by differentiating  
the representation of ${\cal{G}} [ \bd{h} ]$
as a trace of a time-ordered exponential over the physical Hilbert space of the
spin systen. 
For recent applications of this spin FRG approach see
Refs.~[\onlinecite{Tarasevych18,Goll19,Goll20,Tarasevych21,Tarasevych22,Rueckriegel22}]; in particular, this approach 
has been used to calculate the dynamic structure factor of  Heisenberg magnets at infinite temperature \cite{Tarasevych21}, to investigate 
the critical spin dynamics of Heisenberg 
ferromagnets \cite{Tarasevych22}, and  to study
dimerized spin systems \cite{Rueckriegel22}.

In this work we use the spin FRG to investigate the
phase diagram and the critical temperature of a
frustrated quantum spin model in three dimensions.
Specifically, we consider  the Hamiltonian of the  $J_1 J_2J_3$ quantum Heisenberg model
\begin{equation}
{\cal H} =  
  J_1 \sum_{ \langle ij \rangle_1 }  {\bd{S}}_i \cdot {\bd{S}}_j  
+ J_2 \sum_{ \langle ij \rangle_2 }  {\bd{S}}_i \cdot {\bd{S}}_j
+ J_3 \sum_{ \langle ij \rangle_3 }  {\bd{S}}_i \cdot {\bd{S}}_j ,
\label{eq:hamiltonian}
\end{equation}
where the spin $S$ operators $\bd{S}_i$ are
localized at the sites $\bd{R}_i$
of a simple cubic lattice and
$\langle ij \rangle_n$ denotes summation over all distinct pairs of $n$-th nearest neighbor spins. 
This model has recently been used as a benchmark to test 
the accuracy of different implementations
of the 
pseudofermion FRG \cite{Iqbal16,Kiese22,Ritter22}.
In this work we will apply our spin FRG approach to
the  $J_1 J_2 J_3$ model and compare the results to the pseudofermion FRG.
Our main result is that the simplest possible truncation of the spin FRG
produces results which are consistent with the numerically more expensive pseudofermion FRG.
Moreover, 
using more sophisticated truncations of the spin FRG flow equations, 
we can obtain renormalization group fixed points, 
improve our estimates for the critical temperatures, 
and calculate the spin dynamics.
 
The rest of this work is organized as follows: In Sec.~\ref{sec:classical}
we briefly discuss the classical zero temperature phase 
diagram of our model.
In Sec.~\ref{sec:static}, 
we investigate the effect of classical spin fluctuations 
on the critical temperature
using simple static approximations of the spin FRG flow equations,
with and without taking into account the renormalization of the four-spin interaction.
We improve our static estimates in Sec.~\ref{sec:dyn} by including the effect of
dynamic spin fluctuations due to the quantum dynamics of the spin operators.
In Sec.~\ref{sec:summary} we summarize our main results and 
point out necessary modifications of our approach to deal with
strongly frustrated spin systems.
Finally, in four appendices we give additional technical details: 
in Appendix~\ref{app:static} we write down FRG flow equations in static approximation, 
Appendix~\ref{app:fixed} provides a discussion of the renormalization group flow and fixed point 
in the static approximation, 
while in Appendix~\ref{app:single_spin} we derive
initial conditions for the relevant dynamical five and six-point vertices.
The last Appendix~\ref{app:integral} investigates an alternative parametrization of the dynamic spin fluctuations which serves as an input for the FRG calculation of the thermodynamics and 
static correlation function.

\section{Classical phase diagram}
 \label{sec:classical}

\begin{figure}
\centering
\includegraphics[width=\linewidth]{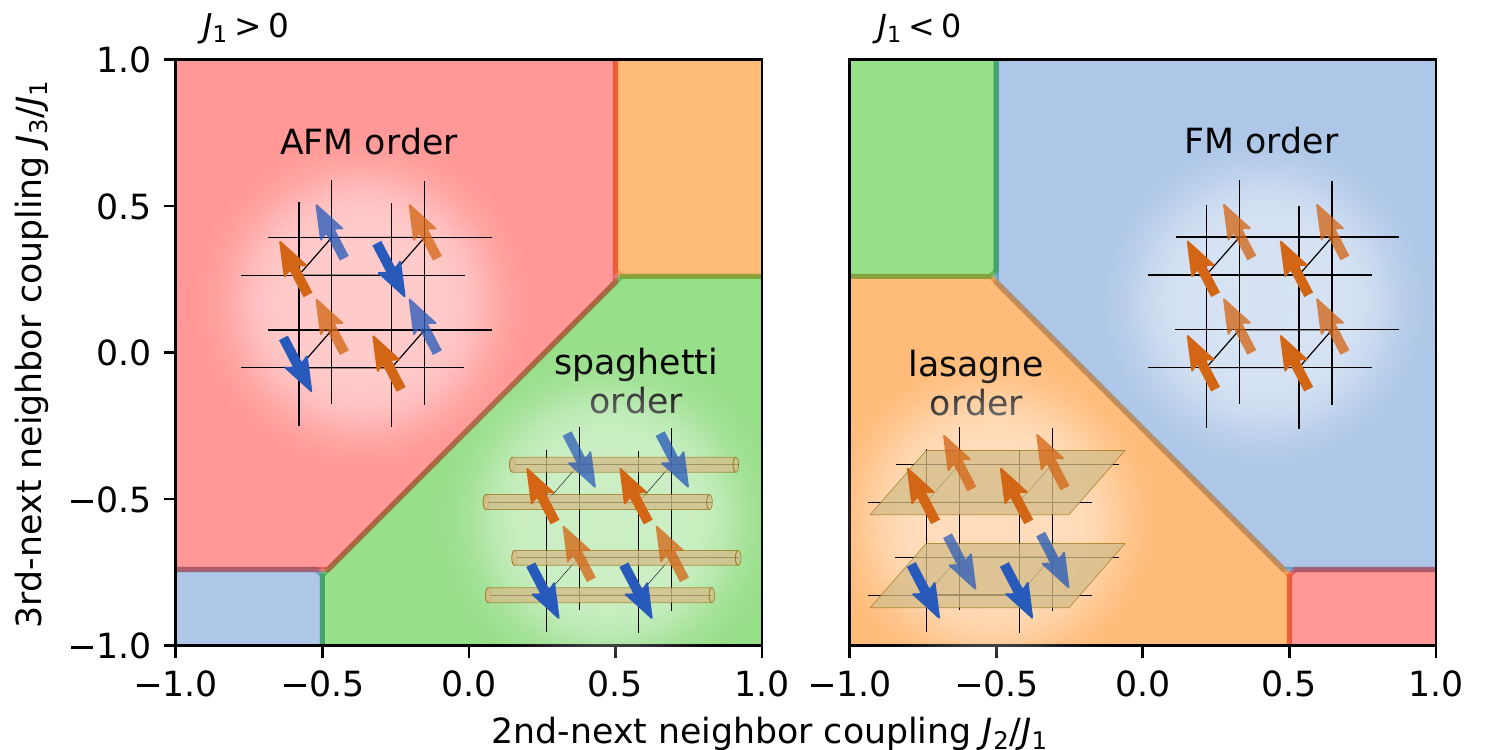}
\caption{%
Classical ground state phase diagram of the $J_1 J_2 J_3$-model on a simple cubic lattice
for $J_1 > 0$ (left panel) and $J_1 < 0$ (right panel).
Depending on the values of $J_2 / J_1$ and $J_3 / J_1$ one of the following four
states has the lowest energy: 
antiferromagnet (AF) with ordering wavevector $\bd{Q} = \bd{R} = ( \pi , \pi , \pi )$;
striped antiferromagnet SP (``spaghetti order'') with $\bd{Q} = \bd{M} = ( 0, \pi , \pi )$;
layered antiferromaget LA (``lasagne order'')  with $\bd{Q} = \bd{X} = ( 0,0, \pi )$; 
and ferromagnet (F)  with $\bd{Q} = \bd{\Gamma} = ( 0,0,0)$.
}
\label{fig:phasediagram123}
\end{figure}

In the classical limit where the spin operators are
replaced by three-component vectors with length $S$
the ground state phase diagram can be obtained by minimizing the 
classical Hamiltonian subject to the constraints $ \bd{S}_i^2  = S^2$.
The classical ground state energy of the
 $J_1 J_2 J_3$ model \eqref{eq:hamiltonian} on a cubic lattice
can then be written as
\begin{equation} 
E_0 = \frac{ N }{ 2 } J_{ \bd{Q} } S^2 ,
\end{equation}
where $\bd{Q}$ is the ordering wavevector
and $N$ is the total number of lattice sites.
For a simple cubic lattice
the Fourier transform of the exchange couplings is 
\begin{align}
J_{\bd{k}} 
& = 6 J_1 \gamma_{\bd{k}}^{(1)} + 12 J_2 \gamma^{(2)}_{\bd{k}} + 8 J_3 \gamma_{\bd{k}}^{(3)},
\label{eq:Jkdef}
\end{align}
where  we have introduced the normalized form factors
\begin{subequations}
\begin{align}
\gamma_{\bd{k}}^{(1)} 
& = \frac{1}{3}  \left( 
\cos k_x + \cos k_y + \cos k_z 
\right) ,
\\
\gamma_{\bd{k}}^{(2)} 
& = \frac{1}{3} \left(
\cos k_x \cos k_y + \cos k_y \cos k_z + \cos k_z \cos k_x  
\right) ,
\\
\gamma_{\bd{k}}^{(3)} 
& = \cos k_x \cos k_y \cos k_z .
\end{align}
\end{subequations}
We measure  wavevectors in units of the inverse
lattice spacing.
According to Ref.~[\onlinecite{Iqbal16}],
in the classical limit one of the following four ground states is realized in the
$J_1 J_2 J_3$ model on a cubic lattice:
antiferromagnet (AF) with ordering wavevector 
$\bd{Q} = \bd{R} = ( \pi , \pi , \pi )$; 
striped AF with $\bd{Q} = \bd{M} = (0, \pi , \pi )$ 
(we call this ``spaghetti order''); 
layered AF with $\bd{Q} = \bd{X} = ( 0,0, \pi )$ (``lasagne order''); 
ferromagnet with $\bd{Q} = \bd{\Gamma} = (0,0,0)$.
From the condition that the physical ground state minimizes $E_0$ 
we obtain the
classical zero temperature phase diagram shown in Fig.~\ref{fig:phasediagram123}.

At finite temperature $T$ the boundaries between the paramagnetic phase and the magnetically ordered phases can be obtained from the momentum-dependent static susceptibility $G ( \bd{k} )$. Assuming that the 
phase transition to the magnetically ordered phase is continuous, we can determine the phase boundaries from the condition that $G ( \bd{Q} )$ diverges at the transition to a magnetically ordered state with ordering wavevector $\bd{Q}$. 
Within a simple mean-field approximation (see below) the static susceptibility is
\begin{equation}
G_0 ( \bd{k} ) = \frac{1}{ J_{\bd{k}} + T / b_1 } ,
\label{eq:G0def}
\end{equation}
where the Fourier transform $J_{\bd{k} }$ of the exchange coupling 
is given in Eq.~(\ref{eq:Jkdef}), and 
\begin{equation}
b_1  =  \frac{ (2S+1)^2 -1}{12} = \frac{ S (S+1) }{3}
\label{eq:b1def}
\end{equation}
is the first coefficient in the expansion of the Brillouin function 
\begin{align}
b(y) 
& =  
\left( S + \frac{1}{2} \right) 
\coth \left[ \left( S + \frac{1}{2} \right) y \right] - \frac{1}{2} \coth \left( \frac{y}{2} \right)
\nonumber
\\
& =
b_1 y + \frac{b_3}{3!} y^3
+ \frac{b_5}{5!} y^5
+ {\cal{O}} \left( y^7 \right) .
 \label{eq:bydef}
\end{align}
In this approximation the critical temperature 
for a  transition to a state with magnetic ordering wavevector
$\bd{Q}$ is given by the mean-field result
 \begin{equation}
T_c^{\rm MF} = - b_1 J_{\bd{Q}} .
\end{equation}
Because $J_{\bd{Q}}$ is the global minimum of $J_{\bd{k}}$,  
the magnetic order in the classical ground state
can  also be identified with the state with the highest critical temperature.

\begin{figure}
\centering
\includegraphics[width=\columnwidth]{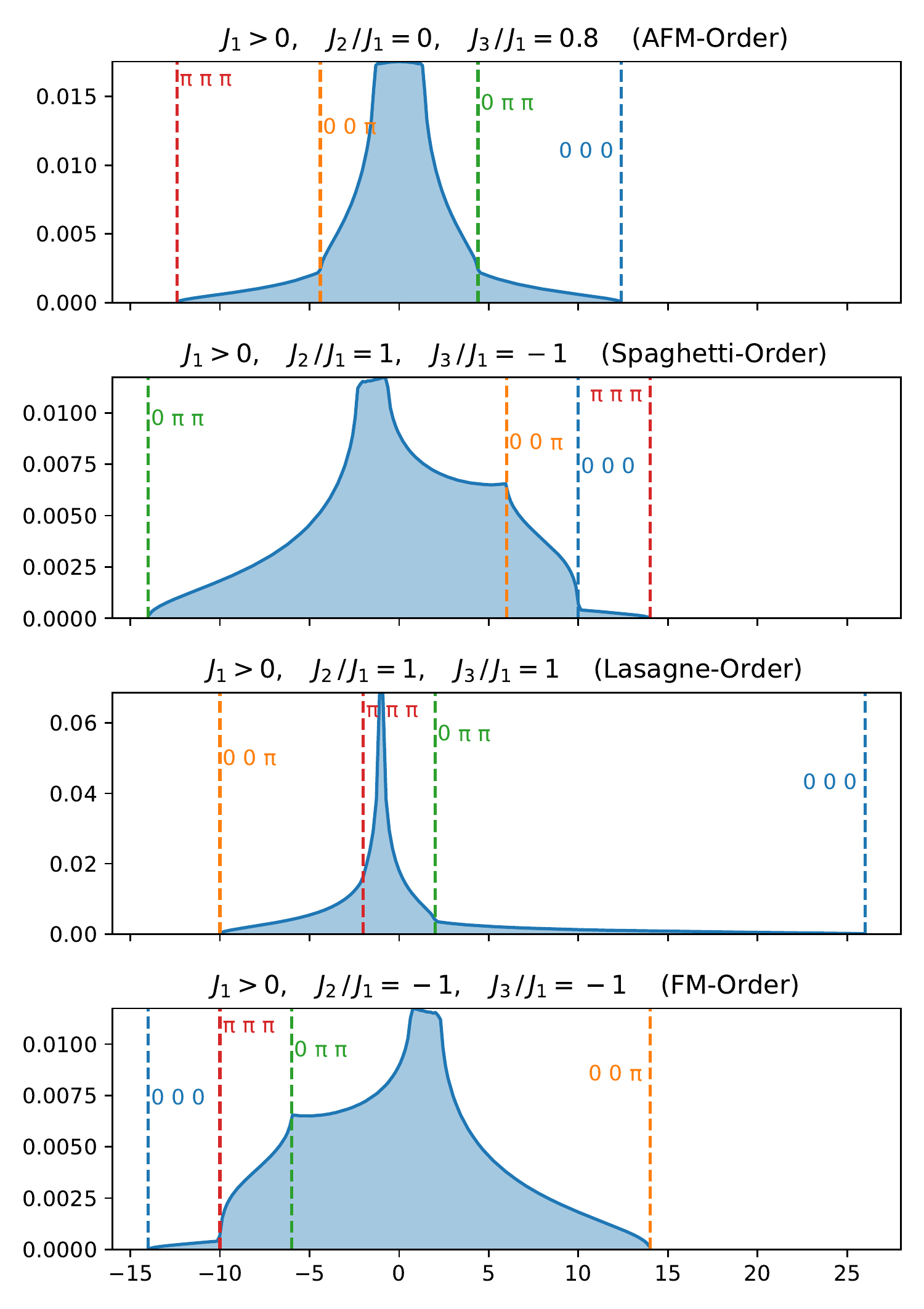}
\caption{
Dimensionless density of states $ J_1 \nu( \epsilon ) $ 
as a function of $\epsilon/J_1$,
at exemplifying values of the coupling strengths $J_2/J_1$ and $J_3/J_1$ corresponding to 
weak frustration.}
\label{fig:dos-examples}
\end{figure}

\begin{figure}
\centering
\includegraphics[width=\columnwidth]{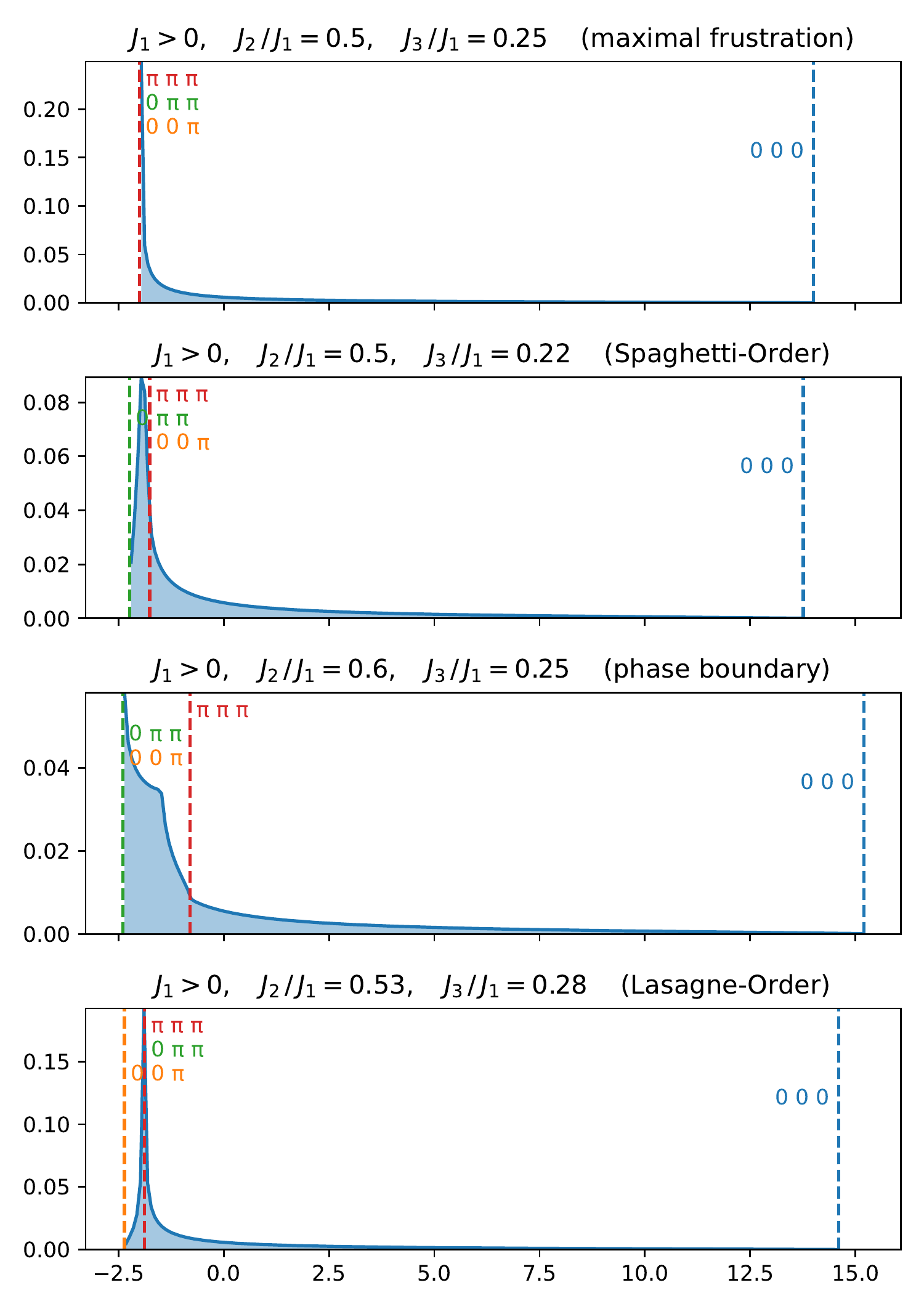}
\caption{Dimensionless density of states $ J_1 \nu( \epsilon ) $ 
as a function of $\epsilon/J_1$ near the tri-border point 
$J_1>0$, $J_2/J_1=1/2$, $J_3/J_1=1/4$; in this case the frustration is strong,
in contrast to the parameter regime shown in Fig.~\ref{fig:dos-examples}.}
\label{fig:dos-benchmark}
\end{figure}

As a quantitative measure for the degree of frustration in the system, 
it is useful to consider the energy-dependent density of states
\begin{equation}
\nu ( \epsilon ) = \frac{1}{N} \sum_{\bd{k}} \delta ( \epsilon - J_{\bd{k} } ) ,
\end{equation}
where the momentum sum is over the first Brillouin zone.
In Figs.~\ref{fig:dos-examples} and \ref{fig:dos-benchmark} we show a numerical evaluation of
$\nu ( \epsilon )$ for $N \rightarrow \infty$ for
representative values of the exchange couplings.
The momentum of the state at the lower band edge determines the
magnetic order in the classical ground state.
If the ground state is separated by a large energy interval
from the other candidate states the system is weakly frustrated,
see Fig.~\ref{fig:dos-examples}.
The corresponding points in the phase diagram shown in
Fig.~\ref{fig:phasediagram123} 
are far away from any of the phase boundaries.
On the other hand, 
if the exchange couplings are chosen such that
the corresponding points in the phase diagram 
are at or close to phase boundaries,
the state at the lower edge of the band is energetically very close
to the other candidate states as illustrated in Fig.~\ref{fig:dos-benchmark}.
Hence, the
frustration is strong in these cases. 
Additionally, 
one can see that
competing states have an enhanced density of states. 
In the extreme cases where several ground states are actually degenerate,
we expect reduced critical temperatures and possibly spin-liquid behavior.
In the following section we will investigate this possibility 
within a simple static truncation of the spin FRG flow equations.

\section{Static spin FRG}

\label{sec:static}

To go beyond the mean-field approximation, we use the spin FRG approach 
proposed in Ref.~[\onlinecite{Krieg19}] and further developed in
Refs.~[\onlinecite{Tarasevych18,Goll19,Goll20,Tarasevych21,Tarasevych22,Rueckriegel22}]. 
For our purpose the hybrid approach developed
in Ref.~[\onlinecite{Tarasevych21}] (see also Appendix~B of Ref.~[\onlinecite{Rueckriegel22}]) is most convenient.
Let us briefly outline the main ideas of this implementation 
of the spin FRG.
First of all, we replace the bare exchange coupling $J_{\bd{k}}$ 
by a deformed coupling $J_{\Lambda} ( \bd{k} )$
parametrized by a continuous scale parameter $\Lambda \in [ 0 , 1 ]$. 
The deformation scheme is chosen such that the
exchange coupling $J_{\Lambda=0} ( \bd{k} )$ 
is sufficiently simple so that the correlation functions of the deformed system can be calculated exactly,
whereas $J_{\Lambda =1} ( \bd{k} ) = J_{ \bd{k} }$ corresponds to the undeformed system.
In this work we use two different deformation schemes:
a simple interaction-switch \cite{Krieg19},
\begin{equation}
J_\Lambda ( \bd{k} ) = \Lambda  J_{ \bd{k} } ,
\label{eq:switch}
\end{equation}
that linearly switches on the exchange couplings,
and a Litim regulator \cite{Litim01, Rancon11a, Rancon11b, Rancon12b, Rancon14, Rueckriegel22}
\begin{align}
J_\Lambda ( \bd{k} ) 
= {} &
J_{ \bd{k} } - 
\Theta\left( J_{ \bd{k} } \right) \left( J_{ \bd{k} } - J_{\rm max } \Lambda \right)
\Theta\left( J_{ \bd{k} } - J_{\rm max } \Lambda \right) 
\nonumber\\
& + 
\Theta\left( - J_{ \bd{k} } \right) \left( - J_{ \bd{k} } + J_{\rm min } \Lambda \right)
\Theta\left( - J_{ \bd{k} } + J_{\rm min } \Lambda \right) 
\label{eq:Litim}
\end{align}
that gradually increases the bandwidth of the exchange couplings,
where $ J_{ {\rm max} ({\rm min}) } $ are the maximum (minimum) value of $J_{ \bd{k} }$.
In both schemes we have $J_{\Lambda =0} ( \bd{k} ) = 0$,
such that the spin correlations at $\Lambda = 0 $  
are site-diagonal and are determined by the dynamics of a single spin.
The corresponding time-ordered correlation functions are highly non-trivial 
and can be obtained diagrammatically by means of the
generalized Wick-theorem for spin operators 
developed by Vaks, Larkin, and Pikin \cite{Vaks68,Vaks68b}, see also
see also Ref.~[\onlinecite{Izyumov88}]. Note that recently \cite{Goll19}
a purely algebraic form of the generalized Wick theorem for spin operators has been  
derived which does not rely on the rather complicated diagrammatic rules
introduced in Refs.~[\onlinecite{Vaks68,Vaks68b,Izyumov88}].
As shown in Ref.~[\onlinecite{Krieg19}],
the deformed generating functional  of the 
imaginary time-ordered spin correlation function satisfies a formally exact FRG flow equation. 
It follows that the subtracted Legendre 
transform $\Gamma_{\Lambda} [ \bd{M} ]$ of ${\cal{G}}_{\Lambda} [ \bd{h} ]$
satisfies the usual Wetterich equation \cite{Wetterich93}.
Unfortunately, at the initial scale $\Lambda =0$ 
where the deformed exchange coupling vanishes, 
the Legendre transform $\Gamma_{0} [ \bd{M} ]$ 
does not exist due to the vanishing of dynamic two-spin correlations of an isolated spin \cite{Rancon14,Krieg19}.
For this reason the lattice FRG proposed by Machado and Dupuis \cite{Machado10}
for classical spin systems cannot be straightforwardly generalized to quantum spin systems.
In Ref.~[\onlinecite{Tarasevych21}] we have solved  this problem by
decomposing spin fluctuations into static and dynamic components
and performing a Legendre transformation only in the static sector. 
Dynamic fluctuations are treated differently by 
working with ``hybrid vertices'' which are 
interaction-irreducible in the dynamic sector.
These vertices are generated by  the 
hybrid functional $\Gamma_{\Lambda} [ \bd{m} , \bd{\eta} ]$
introduced in  Ref.~[\onlinecite{Tarasevych21}],
which depends on the static (classical) magnetization $\bd{m}$ 
and on a dynamic auxiliary field $\bd{\eta}$ that 
can be interpreted as the frequency-dependent part of an internal 
magnetic field generated by the exchange interaction.

In the paramagnetic phase,
the scale-dependent static spin susceptibility
can then be written as
\begin{equation}
G_{\Lambda} ( \bd{k} ) = \frac{1}{ J_{\Lambda} ( \bd{k} ) + \Sigma_{\Lambda} ( \bd{k} ) } ,
\label{eq:Gstatdef}
\end{equation}
where $\Sigma_\Lambda ( \bd{k} )$ is the scale-dependent spin self-energy
with initial condition 
\begin{equation} \label{eq:Sigma_0}
\Sigma_0 ( \bd{k} ) = T / b_1.
\end{equation}
The mean-field result \eqref{eq:G0def}
corresponds to neglecting the flow of this self-energy.
Assuming that a possible phase transition to a magnetically ordered state is continuous,
the critical temperature can be determined from the condition that 
the spin susceptibility \eqref{eq:Gstatdef}
at the ordering wavevector $\bd{k} = \bd{Q}$
diverges at the end of the flow.

To go beyond the mean-field approximation,
let us now consider
the flow of the static spin self-energy
$\Sigma_{\Lambda} ( \bd{k} )$.
In this section we use the static approximation, 
where all vertices involving external legs with finite frequencies 
are  neglected. Formally, this amounts to
setting $\Gamma_{\Lambda} [ \bd{m} , \bd{\eta} ]
 \approx \Gamma_{\Lambda} [ \bd{m} , \bd{\eta} =0 ]$.
Given the fact that finite-temperature critical behavior 
is completely determined by classical fluctuations,
we expect that the static approximation is sufficient to obtain 
the fixed point of the renormalization group flow
associated with a finite temperature phase transition. 
The spin self-energy defined via Eq.~(\ref{eq:Gstatdef})
then satisfies the flow equation
\begin{equation}
\partial_{\Lambda} \Sigma_{\Lambda} ( \bd{k} ) =
\frac{T}{N} \sum_{\bd{q}} \dot{G}_{\Lambda} ( \bd{q} ) 
\Gamma^{(4)}_{\Lambda} ( - \bd{q} ,  \bd{q} , - \bd{k} , \bd{k} ) ,
\label{eq:Sigmaflow}
\end{equation}
where the single-scale propagator is defined by
\begin{equation}
\dot{G}_{\Lambda} ( \bd{k} ) = - G^2_{\Lambda} ( \bd{k} )
\partial_{\Lambda} J_{\Lambda} ( \bd{k} )
\end{equation}
and the irreducible four-point vertex
$\Gamma^{(4)}_{\Lambda} ( - \bd{q} ,  \bd{q} , - \bd{k} , \bd{k} )$ 
describes the interaction between four spins in the static limit. Note that the three-point vertex $\Gamma^{(3)}_{\Lambda}$ vanishes because we consider the paramagnetic phase
where there is no spontaneous magnetization.
With our deformation scheme where the deformed exchange interaction initially vanishes,
the initial value of the four-point vertex is determined by the irreducible part of 
a rotationally invariant combination of
four-spin correlation functions of an isolated spin
in the static limit \cite{Goll19, Tarasevych21}. 
In Appendix~\ref{app:static} we show that
\begin{equation}
\Gamma^{(4)}_0 ( - \bd{q} ,  \bd{q} , -\bd{k} , \bd{k} ) 
= \frac{5}{6} U_0 , \; \; \; \; \; \;
U_0 = - T \frac{ b_3 }{ b_1^4} > 0 ,
\label{eq:Gamma4init}
\end{equation}
where
\begin{equation}
b_3 
= - \frac{ (2S+1)^4 -1}{120} 
= - \frac{6}{5} b_1 \left( b_1 + \frac{1}{6} \right) ,
\label{eq:b3def}
\end{equation}  
is the third order coefficient in the Taylor expansion \eqref{eq:bydef}
of the Brillouin function.


\subsection{Level-1 truncation}

\label{sec:level-1}

In the simplest level-1 truncation \cite{Metzner12} 
we approximate the four-point vertex by its initial value
given in Eq.~\eqref{eq:Gamma4init}.
A similar level-1 truncation of the hierarchy of FRG flow equations
has been used to calculate the renormalization of impurity potentials 
in mesoscopic Luttinger liquids \cite{Meden02,Metzner12}.
The spin self-energy $\Sigma_{\Lambda} ( \bd{k} ) = \Sigma_{\Lambda}$ 
is then independent of the momentum $\bd{k}$ and
its flow equation \eqref{eq:Sigmaflow} reduces to
\begin{equation}
\partial_{\Lambda} \Sigma_{\Lambda}  =
- \frac{ 5 }{ 6 } U_0 \frac{T}{N} \sum_{\bd{q}} 
\frac{ \partial_\Lambda J_\Lambda ( \bd{q} ) 
}{ \left[ J_\Lambda ( \bd{q} ) + \Sigma_{\Lambda} \right]^2 } ,
\label{eq:Sigmaflow_1}
\end{equation}
which can be straightforwardly integrated numerically 
with the initial condition \eqref{eq:Sigma_0}.

\begin{figure}[tb]
\centering
\includegraphics[width=\linewidth]{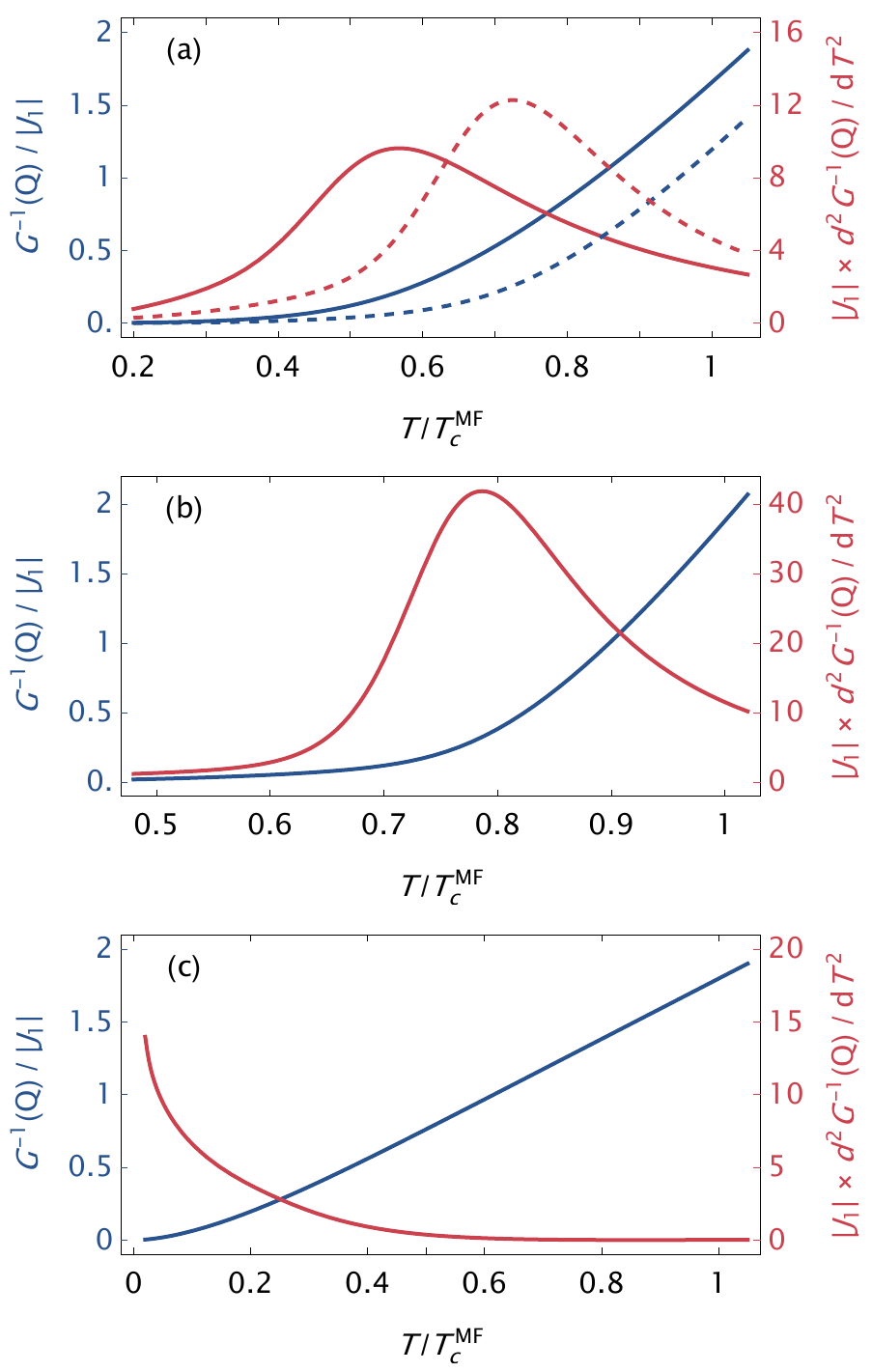}
\caption{
Temperature dependence of the inverse spin susceptibility $G^{-1} ( \bd{Q} )$ 
and its second derivative in level-1 truncation.
(a) Nearest neighbor Heisenberg magnet,
$J_1 \neq 0$, $J_2 = 0 = J_3$,
for $S = 1 / 2$ (solid lines) 
and in the classical limit $S \to \infty$ (dashed lines)  \cite{footnote_classical},
calculated with the Litim deformation scheme \eqref{eq:Litim}.
(b) $J_1 > 0$, $J_2 = 0$, $J_3 / J_1 = 0.8$,
and $S = 1 / 2$,
calculated with the Litim deformation scheme \eqref{eq:Litim}.
(c) $J_1 > 0$, $J_2 / J_1 = 0.6$, $J_3 / J_1 = 0.25$,
and $S = 1 / 2$,
calculated with the interaction-switch deformation scheme \eqref{eq:switch}.
}
\label{fig:level1}
\end{figure}

In Fig.~\ref{fig:level1} 
we show our numerical results for the inverse susceptibility 
$G_{ \Lambda = 1 }^{ - 1 } ( \bd{Q} ) = G^{ - 1 } ( \bd{Q} ) $ 
at the end of the FRG flow
as a function of the dimensionless temperature $T / T_c^{\rm MF}$
for three different sets of exchange couplings.
Note that for the nearest neighbor exchange ($J_1 \neq 0$, $J_2 = J_3 =0$) displayed in Fig.~\ref{fig:level1} (a),
our flow equation \eqref{eq:Sigmaflow_1}
is independent of the sign of the nearest neighbor coupling $J_1$.
Therefore we obtain the same temperature dependence of the susceptibility
for ferro- and antiferromagnets,
which is only correct in the classical $S \to \infty$ limit \cite{footnote_classical,Oitmaa04}.
Remedying this for finite $S$ requires the inclusion of dynamical quantum fluctuations,
which will be addressed in Sec.~\ref{sec:dyn}.
Obviously,
within the level-1 truncation used in this section the inverse susceptibility
approaches zero only asymptotically for $T \to 0$,
implying a paramagnetic state at finite temperature.
This is of course an artifact of the level-1 truncation,
which neglects the renormalization of the four-point vertex.
We show in Appendix \ref{app:fixed}
that a true fixed point of the renormalization group flow can only be obtained 
if the flow of the four-point vertex is taken into account.
Nevertheless,
the existence of a kink in the temperature dependence of the inverse susceptibility
shown in Figs.~\ref{fig:level1} (a) and (b)
suggests that a large part of the renormalization group flow in the level-1 truncation 
still ``feels'' the influence of an underlying critical fixed point
for these parameters.
We therefore estimate the critical temperature by the position of the associated maximum in the second derivative
of the inverse susceptibility with respect to temperature.
In Table~\ref{table_1},
we have collected the critical temperatures obtained in this manner,
using both deformation schemes discussed above,
and compare them to the accepted benchmark values 
from quantum Monte-Carlo simulations \cite{Sandvik98,Troyer04,Iqbal16}
and high-temperature expansions \cite{Oitmaa04}.
\begin{table}
\centering
\begin{tabular}{ c c c c c c c c c c c }
\hline
\hline
&&&&& \multicolumn{3}{c}{$T_c / T_c^{\rm MF}$} && \multicolumn{2}{c}{rel. error / $\%$} \\
\cline{6-8}
\cline{10-11}
$S$				&&	 $J_1$		&	 $J_3 / J_1$	&&	switch	&	Litim	&	benchmark	&&	switch	&	Litim	\\
\hline
$1/2$ 		&&	$<0$			&	0		&&	0.651		&	0.568	&	0.559		&&	16.5	&	1.6		\\
$1/2$ 		&&	$>0$			&	0		&&	0.651		&	0.568	&	0.629		&&	3.5		&	9.7		\\
$1$ 			&&	$<0$			&	0		&&	0.726		&	0.668	&	0.650		&&	11.7	&	2.8		\\
$1$ 			&&	$>0$			&	0		&&	0.726		&	0.668	&	0.684		&&	6.1		&	2.3		\\
$3/2$ 		&&	$<0$			&	0		&&	0.745		&	0.695	&	0.685		&&	8.8		&	1.5		\\
$3/2$ 		&&	$>0$			&	0		&&	0.745		&	0.695	&	0.702		&&	6.1		&	1.0		\\
$1/2$ 		&&	$>0$			&	0.2	&&	0.746		&	0.701	&	0.722		&&	3.3		&	2.9		\\
$1/2$ 		&&	$>0$			&	0.4	&&	0.782		&	0.753	&	0.768		&&	1.8		&	2.0		\\
$1/2$ 		&&	$>0$			&	0.6	&&	0.800		&	0.776	&	0.794		&&	0.8		&	2.3		\\
$1/2$ 		&&	$>0$			&	0.8	&&	0.807		&	0.787	&	0.808		&&	0.1		&	2.6		\\
$\infty$	&&	$\neq 0$	&	0		&&	0.766		&	0.725	&	0.722		&&	6.1		&	0.4		\\
\hline
\hline
\end{tabular}
\caption{
Level-1 critical temperatures for the $J_1 J_3$ model with $J_2 = 0$,
extracted from the maximum of $d^2 G^{-1} ( \bd{Q} ) / d T^2$, 
both with the interaction-switch 
and the Litim deformation scheme,
Eqs.~\eqref{eq:switch} and \eqref{eq:Litim} respectively.
For comparison, we also show the 
accepted benchmark values from quantum Monte-Carlo simulations \cite{Sandvik98,Troyer04,Iqbal16}
and high-temperature expansions \cite{Oitmaa04},
as well as the relative error of the level-1 results.  
Note that the Litim scheme always predicts a lower $T_c$ than the interaction-switch.
}
\label{table_1}
\end{table}
The overall agreement is rather good,
with the largest deviation for the $S=1/2$ nearest neighbor case,
where the static approximation does not distinguish between 
ferro- and antiferromagnetic nearest-neighbor couplings.

Next, let us consider Fig.~\ref{fig:level1} (c) in more detail.
For $J_1 > 0$, $J_2 / J_1 = 0.6$, $J_3 / J_1 = 0.25$,
the classical ground states with
spaghetti-order [$\bd{Q} = ( 0, \pi , \pi )]$ and lasagne-order [$\bd{Q} = ( 0, 0, \pi)$] are degenerate
at the mean-field level;
see Figs.~\ref{fig:phasediagram123} and \ref{fig:dos-benchmark}.
For this specific set of parameters an  older one-loop pseudofermion FRG study \cite{Iqbal16} 
reported evidence for a paramagnetic ground state, 
whereas  a more sophisticated multiloop 
pseudofermion FRG \cite{Ritter22} found that eventually the system exhibits
spaghetti order (striped AF) at low temperatures.
As our spin FRG results do not exhibit any kink as a function of temperature for these parameters,
our calculation suggests  a paramagnetic ground state, 
in agreement with the older pseudofermion FRG results by Iqbal {\it{et al.}} \cite{Iqbal16}.

We conclude that, at least for the three-dimensional $J_1 J_2 J_3$ model 
on a cubic lattice, a simple static level-1 truncation of the spin FRG flow equations 
(where the frequency-dependence of all vertices 
as well as the renormalization of the four-point vertex are neglected)  
gives results for the critical temperature of similar accuracy as
the numerically more expensive multiloop pseudofermion FRG.
On the other hand, the fact that our static level-1 truncation of the spin FRG flow equations
does  not reproduce  the magnetic order found in a recent
multiloop  pseudofermion FRG \cite{Ritter22} in a regime
 where classically the system exhibits 
degenerate ground states
might indicate that
in this  regime  our level-1 truncation
possibly tends to over-estimate the role of spin fluctuations.

\subsection{Level-2 truncation}

\label{sec:level-2}

The absence of a sharp phase transition in the level-1 truncation 
is due to the fact that within this truncation the flow of the four-point vertex is neglected.
As shown in Appendix \ref{app:fixed},
this leads to a runaway flow of the rescaled couplings.
In the parameter regime where the classical ground state is not degenerate, 
it is however straightforward to recover a fixed point within our spin FRG approach
in a level-2 truncation which  takes the renormalization of the four-spin interaction into account. 
Note that within the pseudofermion FRG 
the four-spin interaction is encoded in the fermionic eight-point vertex which is usually neglected \cite{Reuther10,Reuther11,Reuther11a,Buessen16,Iqbal16,Baez2017,Rueck18,Thoenniss20,Kiese22,Ritter22}.
On the other hand, within the pseudo-Majorana FRG \cite{Niggemann2021,Niggemann22} 
the renormalization of the four-spin interaction
can  at least partially be taken into account due to an operator identity relating products involving different 
numbers of Majorana fermions \cite{Mao03,Shnirman03}.

The FRG flow equation for the momentum-dependent
static four-point vertex
$\Gamma^{(4)}_{\Lambda} ( \bd{k}_1 , \bd{k}_2 , \bd{k}_3 , \bd{k}_4 )$ 
that determines the flow of the spin self-energy via
Eq.~(\ref{eq:Sigmaflow}) is given in Appendix~\ref{app:static}.
In practice, additional approximations are necessary.
For simplicity, let us focus here on non-degenerate classical ground states.
Then we can adopt the truncation strategy of Ref.~[\onlinecite{Krieg19}], 
where the critical temperature of the three-dimensional Ising model 
has been obtained to an accuracy of about $1 \%$ 
using a truncation where the momentum-dependence of the renormalized four-point vertex is neglected 
and the six-point vertex is approximated by its initial value. 
A similar truncation strategy for the Heisenberg model leads to the following
flow equation for the renormalized four-point vertex
$\Gamma^{(4)}_{\Lambda} ( \bd{k}_1 , \bd{k}_2 , \bd{k}_3 , \bd{k}_4 ) \approx \frac{5}{6} U_\Lambda$:
\begin{equation}
\partial_{\Lambda} U_{\Lambda} 
= 
\frac{T}{N} \sum_{\bd{q}} \dot{G}_{\Lambda} ( \bd{q} ) 
\left[ \frac{7}{10} V_{0}  - \frac{11}{3} U_{\Lambda}^2 G_{\Lambda} ( \bd{q} ) \right],
\label{eq:Uflow2}
\end{equation}
where
\begin{equation}
V_0
= T \left( 10  \frac{ b_3^2 }{ b_1^7 } - \frac{b_5}{b_1^6 } \right)
\label{eq:V0}
\end{equation}
is the initial value of the longitudinal six-point vertex defined in Appendix~\ref{app:static}.
Here $b_1$ and $b_3$ are defined in Eqs.~\eqref{eq:b1def} and \eqref{eq:b3def} respectively, 
and
\begin{equation}
b_5 = \frac{ (2S+1)^6 -1}{252}
\label{eq:b5def}
\end{equation}
is the fifth-order coefficient in the expansion \eqref{eq:bydef} of the Brillouin function.
The corresponding flow of the static self-energy is obtained by replacing
$U_0 \rightarrow U_{\Lambda}$ in Eq.~\eqref{eq:Sigmaflow_1},
\begin{equation}
\partial_{\Lambda} \Sigma_{\Lambda}  =
- \frac{ 5 }{ 6 } U_\Lambda \frac{T}{N} \sum_{\bd{q}} 
\frac{ \partial_\Lambda J_\Lambda ( \bd{q} ) 
}{ \left[ J_\Lambda ( \bd{q} ) + \Sigma_{\Lambda} \right]^2 } .
\label{eq:Sigmaflow_2}
\end{equation}

\begin{figure}[tb]
\centering
\includegraphics[width=\linewidth]{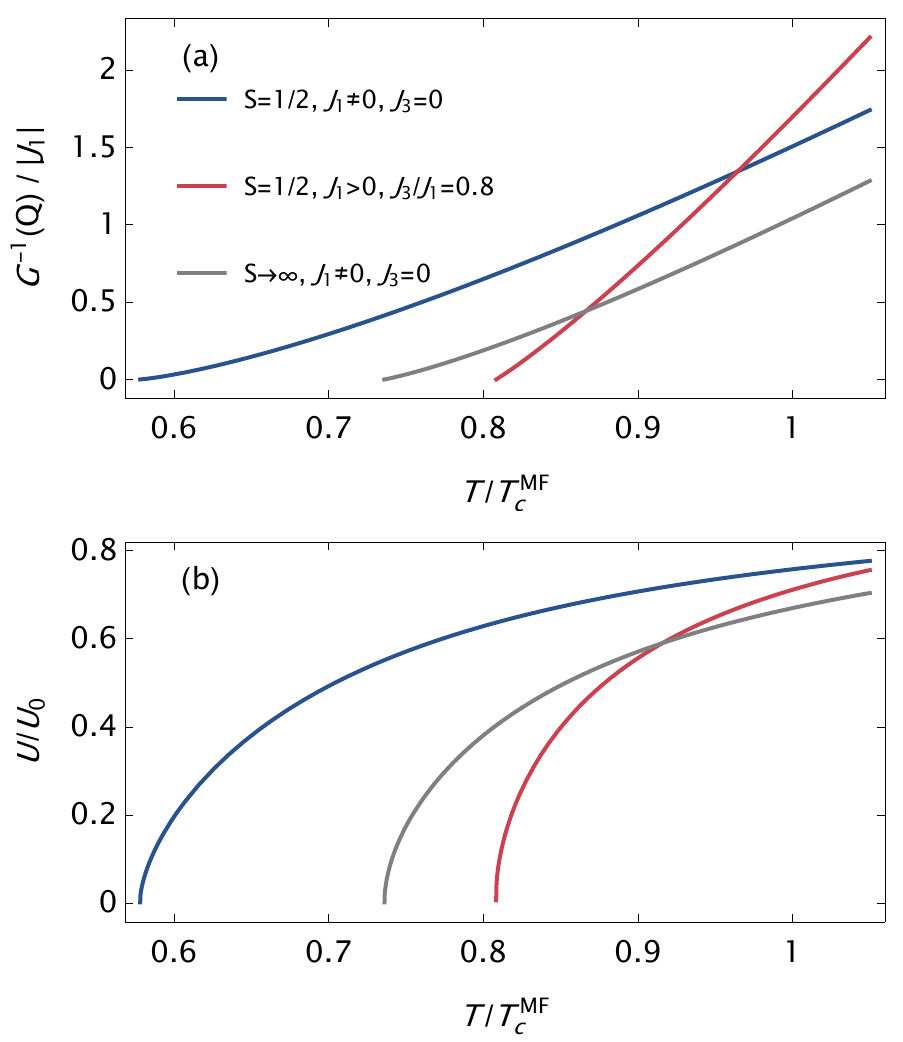}
\caption{
Level-2 temperature dependence of 
(a) the inverse spin susceptibility $G^{-1} ( \bd{Q} )$ and
(b) the four-point vertex $U$,
for the $J_1 J_3$ model with $J_2 = 0$ and
the interaction-switch deformation scheme \eqref{eq:switch}. 
}
\label{fig:level2}
\end{figure}

Our numerical results for the inverse susceptibility
and the four-point vertex $U_{\Lambda = 1 } = U$ at the end of the flow 
are displayed in Fig.~\ref{fig:level2}
as a function of the dimensionless temperature $T / T_c^{\rm MF}$.
In contrast to the level-1 truncation, 
the inverse susceptibility as well as the four-point vertex 
now vanish at a critical temperature $T_c$,
signaling a phase transition.
We explicitly show in Appendix \ref{app:fixed} 
how this phase transition is governed by the Wilson-Fisher fixed point 
for the nearest-neighbor Heisenberg magnet.
The associated values for $T_c$ are listed in Table \ref{table_2}.
\begin{table}
\centering
\begin{tabular}{ c c c c c c c c c c c }
\hline
\hline
&&&&& \multicolumn{3}{c}{$T_c / T_c^{\rm MF}$} && \multicolumn{2}{c}{rel. error / $\%$} \\
\cline{6-8}
\cline{10-11}
$S$				&&	 $J_1$		&	 $J_3 / J_1$	&&	switch	&	Litim	&	benchmark	&&	switch	&	Litim	\\
\hline
$1/2$ 		&&	$<0$			&	0		&&	0.578		&	0.525	&	0.559		&&	3.4		&	6.1		\\
$1/2$ 		&&	$>0$			&	0		&&	0.578		&	0.525	&	0.629		&&	8.1		&	16.5		\\
$1$ 			&&	$<0$			&	0		&&	0.672		&	0.625	&	0.650		&&	3.4		&	3.8		\\
$1$ 			&&	$>0$			&	0		&&	0.672		&	0.625	&	0.684		&&	1.8		&	8.6		\\
$3/2$ 		&&	$<0$			&	0		&&	0.701		&	0.658	&	0.685		&&	2.3		&	3.9		\\
$3/2$ 		&&	$>0$			&	0		&&	0.701		&	0.658	&	0.702		&&	0.1		&	6.3		\\
$1/2$ 		&&	$>0$			&	0.2	&&	0.712		&	0.676	&	0.722		&&	1.4		&		6.4	\\
$1/2$ 		&&	$>0$			&	0.4	&&	0.768		&	0.740	&	0.768		&&	0.0		&		3.7	\\
$1/2$ 		&&	$>0$			&	0.6	&&	0.795		&	0.771	&	0.794		&&	0.1		&	2.9		\\
$1/2$ 		&&	$>0$			&	0.8	&&	0.808		&	0.787	&	0.808		&&	0.0		&	2.6		\\
$\infty$	&&	$\neq 0$	&	0		&&	0.736		&	0.700	&	0.722		&&	1.9		&	3.0		\\
\hline
\hline
\end{tabular}
\caption{
Same as Table \ref{table_1},
but for the level-2 truncation where there is a fixed point in the FRG flow.
}
\label{table_2}
\end{table}
Note the striking agreement with the accepted benchmark values,
in particular for the interaction-switch deformation scheme.
In contrast,
the tendency of the Litim scheme to underestimate the value of $T_c$
even worsens compared to the level-1 results displayed in Table \ref{table_1}. 
A possible explanation for the better performance of the interaction-switch deformation for the calculation of $T_c$ 
is that with this deformation scheme our truncation is
perturbatively controlled in $\Lambda |J ( \bm{k} )|/T$,
which is small  
as long as the deformation parameter $\Lambda$ is sufficiently small or the temperature is large. 
Note furthermore that the interaction-switch deformation does not modify the momentum dependence of $J_\Lambda(\bm{k})$, 
in contrast to the Litim cutoff.
We also remark that the accuracy of the static truncation quickly increases with increasing $S$,
reflecting the decreasing relevance of dynamic fluctuations in this case; see the following Sec. \ref{sec:dyn}.

\section{Including dynamic spin fluctuations}
\label{sec:dyn}

\begin{figure}
\centering
\includegraphics[width=\linewidth]{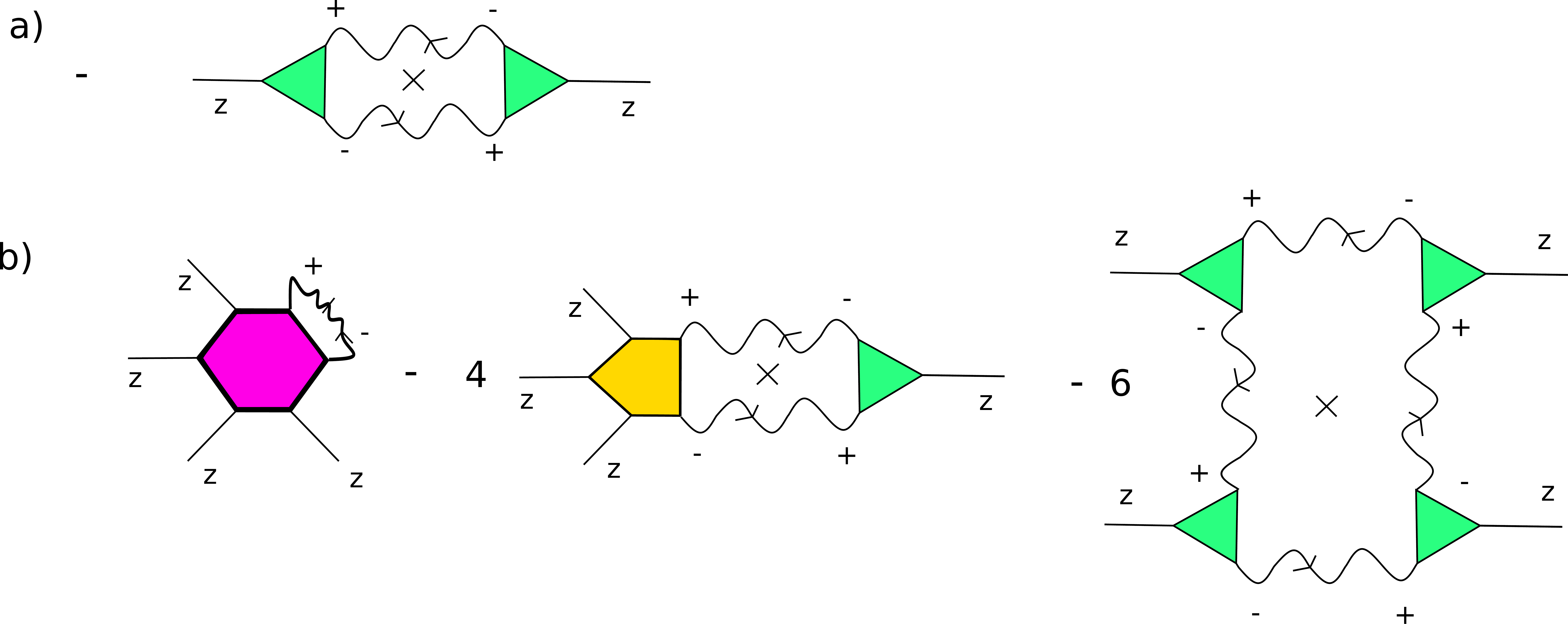}
\caption{
Graphical representation of the frequency-dependent terms 
on the right-hand side of the flow equations \eqref{eq:Sigmaflow3} (a) and \eqref{eq:Uflow3} (b). 
A directed wavy internal line represents the  propagator $F_\Lambda(K)$ at finite frequency. An additional slash means the associated single-scale propagator $\dot{F}_\Lambda(K)$.
The cross inside of each loop means that each propagator is
successively replaced by the corresponding single-scale propagator.
}
\label{fig:diagramsdyn}
\end{figure}

While the static approximation of the previous Sec.~\ref{sec:static}
already yields reasonable results for $T_c$ for various quantum magnets,
it fails to distinguish between the nearest-neighbor ferro- and antiferromagnets at finite $S$.
To remedy this deficiency,
we include also dynamic fluctuations at finite frequencies in this section.
Within the hybrid approach developed in Ref.~[\onlinecite{Tarasevych21}],
such quantum fluctuations are described by
dynamic vertices involving at least one finite-frequency leg 
associated with the auxiliary field $\bd{\eta}$
that represents the dynamic internal exchange field. 
To leading order, these vertices can be approximated by 
their initial values describing the dynamic correlations of an isolated spin.
The leading dynamic modification of the flow of the static spin self-energy 
is then given by the diagram shown in Fig.~\ref{fig:diagramsdyn}~(a), 
where the green triangle represents the 
dynamic chiral three-point vertex \cite{Tarasevych21},
\begin{equation}
\Gamma_0^{z \eta^- \eta^+} ( 0, - \omega , \omega ) = \frac{1}{  i \omega } , 
\; \; \;
\omega \neq 0 ,
\label{eq:Gamma3chiral}
\end{equation}
where the superscripts denote the type of external legs: 
$z$ represents a static fluctuation of the longitudinal magnetic field, while $\eta^{+} = (\eta^{x} + i \eta^{y})/\sqrt{2}$ and $\eta^- = (\eta^{x} - i \eta^{y})/\sqrt{2}$ represent the spherical components of the dynamic exchange field $\bd{\eta}$.
Note that in a Cartesian basis this vertex is determined by the chiral on-site 
expectation value
$ \int_0^{\beta} d \tau_3 \langle S_i^x ( \tau_1 ) S_i^y ( \tau_2 ) S_i^z ( \tau_3 )\rangle$.
For a non-degenerate ground state,
we should then evaluate
the flow of the static self energy $\Sigma_\Lambda$
and the static four-point vertex $\Gamma^{ (4) }_\Lambda$
at the ordering wavevector $\bd{Q}$.
In this case the flow equation \eqref{eq:Sigmaflow_2} for the
static self-energy is replaced by
\begin{align}
\partial_{\Lambda} \Sigma_{\Lambda} 
= {} &    
\frac{5}{6} U_{\Lambda} \frac{T}{N} \sum_{ \bd{q} } \dot{G}_{\Lambda} ( \bd{q} )
\nonumber\\
& 
+ \frac{T}{N} \sum_{ \omega \neq 0 } \sum_{ \bd{q} } \frac{2}{\omega^2}
\dot{F}_{\Lambda } ( \bd{q} , i \omega )   {F}_{\Lambda } ( \bd{q} + \bd{Q} , i \omega ) ,
\label{eq:Sigmaflow3}
\end{align}
where the dynamical propagator $F_{\Lambda} ( K )$ and its single-scale counterpart
$\dot{F}_{\Lambda} ( K )$ are defined by
\begin{subequations}
\begin{align}
F_{\Lambda} ( K ) 
& =   
- \frac{ {G}^{-1}_{\Lambda} (  \bd{k} ) 
}{ 1 + {G}^{-1}_{\Lambda} (  \bd{k} ) \tilde{\Pi}_{\Lambda} ( K ) } ,
\label{eq:Fdef2}
\\
\dot{F}_{\Lambda} ( K )
& = 
-\frac{  \partial_{\Lambda} J_\Lambda ( \bd{k} ) 
}{ [ 1 + {G}^{-1}_{\Lambda} ( \bd{k} ) \tilde{\Pi}_{\Lambda} ( K ) ]^2 } .
\label{eq:etaprop2}
\end{align}
\end{subequations}
Here, $\tilde{\Pi}_\Lambda ( K )$ is the irreducible dynamic 
spin-susceptibility \cite{Tarasevych21}.
Note that in the classical $ S \to \infty $ limit,
the contribution of dynamic spin fluctuations to the flow of $\Sigma_\Lambda$ vanishes
as $1/S^2$ after appropriate rescaling \cite{footnote_classical}.
Hence, 
the classical limit corresponds to the static truncation of the flow equations,
whereas the dynamical terms describe the effect of quantum fluctuations at finite $S$.

For consistency, we should also take the effect of chiral dynamic fluctuations 
in the flow of the four-point vertex into account.
Approximating the relevant higher-order dynamic vertices by their initial values,
we find that the flow equation \eqref{eq:Uflow2}
for the static four-point vertex is modified as follows,
\begin{widetext}
\begin{align}
\partial_{\Lambda} U_{\Lambda} 
= {} &  
\frac{T}{N} \sum_{ \bd{q} } 
\dot{G}_{\Lambda} ( \bd{q} ) \left[ 
\frac{7}{10} V_{0} - \frac{11}{3} U_{\Lambda}^2 G_{\Lambda} ( \bd{q} ) 
\right]
\nonumber
\\
& 
+ \frac{T}{N} \sum_{ \omega \neq 0 } \sum_{ \bd{q} }
\dot{F}_{\Lambda } ( \bd{q} , i \omega )
\Gamma_{0}^{zzzz\eta^- \eta^+} ( 0,0,0,0, - \omega , \omega   ) 
\nonumber\\
& 
- 4 \frac{T}{N} \sum_{ \omega \neq 0 } \sum_{ \bd{q} }
2 \dot{F}_{\Lambda } ( \bd{q} , i \omega ) {F}_{\Lambda } ( \bd{q} + \bd{Q} , i \omega ) 
\Gamma_{0}^{zzz\eta^- \eta^+} ( 0,0,0, - \omega , \omega   ) 
\Gamma_0^{z \eta^- \eta^+ } ( 0, - \omega ,  \omega )
\nonumber\\
& 
- 6 \frac{T}{N} \sum_{ \omega \neq 0 } \sum_{ \bd{q} } 
4 \dot{F}_{\Lambda } ( \bd{q} , i \omega )   
{F}_{\Lambda } ( \bd{q}  , i \omega ) {F}^2_{\Lambda } ( \bd{q} + \bd{Q} , i \omega )
 \left[ \Gamma_0^{z \eta^- \eta^+} ( 0, - \omega , \omega ) \right]^4 .
\label{eq:Uflow3}
\end{align}
\end{widetext}
The frequency-dependent  terms in  Eq.~\eqref{eq:Uflow3}
are shown diagrammatically in Fig.~\ref{fig:diagramsdyn}~(b).
As in Sec.~\ref{sec:static}, 
there is no momentum transfer in the static part of the flow
because $\bd{Q} \pm \bd{Q}$ is always a reciprocal lattice vector.
For the explicit evaluation of the flow equation \eqref{eq:Uflow3} 
of the static four-point vertex,
we also need the initial values of the higher-order dynamic vertices.
As shown in Appendix \ref{app:single_spin} the relevant initial values of the higher-order vertices are
\begin{align}
\Gamma_{0}^{zzz\eta^- \eta^+} ( 0,0,0, - \omega , \omega   ) 
& 
= 0 ,
\label{eq:5point}
\\
\Gamma_{0}^{zzzz\eta^- \eta^+} ( 0,0,0,0, - \omega , \omega   ) 
&
= \frac{ 4  b_3  }{ \beta b_1^4 \omega^2} .
\label{eq:6point}
\end{align}
Also note that all four-point vertices with two legs at finite frequency, e.g. $\Gamma^{zz\eta^-\eta^+}_\Lambda(0,0,-\omega,\omega)$, are initially zero \cite{Tarasevych21} and therefore diagrams containing them do not contribute to the flow equations.
To close the system of flow equations \eqref{eq:Sigmaflow3} and \eqref{eq:Uflow3},
we still need an expression for the dynamic irreducible susceptibility
$\tilde{\Pi}_{\Lambda} ( K )$.
In principle,
we can also write down a flow equation for this function which has been derived in Ref.~[\onlinecite{Tarasevych21}].
In this work,
we instead opt to use the leading term in the high-temperature expansion,
\begin{align}
\tilde{\Pi}_{\Lambda} ( \bd{k} , i \omega ) 
= {} &
\frac{2 b_1^2 }{ T \omega^2 } \int_{\bd{q}} J_{\Lambda} ( \bd{q} ) \left[ 
J_\Lambda ( {\bd{q}} )  - J_{\Lambda} ( {\bd{q} + \bd{k}} ) 
\right]
\nonumber\\
&
+ {\cal{O}} \left( J_\Lambda^3 / T^4 \right) .
\label{eq:Pi_high-T}
\end{align}
which, for instance, can be obtained by iterating the flow of $\tilde \Pi_\Lambda(\bd{k}, i \omega)$ up to second order in $J_\Lambda$ \cite{Kriegthesis}.
Such a high-temperature approximation is of course only valid for 
$ T \gg | J_\Lambda ( {\bd{q}} ) | $.
As $ T_c \sim | J_{ \bd{Q} } | $,
we thus expect that 
this approximation always breaks down 
in the vicinity of the phase transition.
However,
the situation is not as bad as it seems 
because \emph{during} the flow
we actually have to compare $T$ with
the \emph{deformed} exchange interaction $ J_\Lambda ( {\bd{q}} ) $.
Since this deformed coupling only gradually increases from zero
to its physical value, 
the high-temperature expansion is valid 
for most of the flow.
Only in the final stage of the flow, 
for $\Lambda \to 1$, 
corrections to the high-temperature approximation \eqref{eq:Pi_high-T}
can become important for $ T \sim T_c $,
which we neglect.
In Appendix \ref{app:integral},
we discuss a more sophisticated ansatz for $ \tilde{\Pi} ( K ) $
that is based on a solution of the flow equation in the high-temperature limit 
\cite{Tarasevych21}.

An advantage of the high-temperature approximation \eqref{eq:Pi_high-T}
is that we can use the calculus of residues to explicitly evaluate all Matsubara sums 
in the flow equations \eqref{eq:Sigmaflow3} and \eqref{eq:Uflow3}.
To that end, we set
\begin{equation}
G_\Lambda^{ - 1 } ( \bd{k} ) \tilde{\Pi}_{\Lambda} ( \bd{k} , i \omega ) =
\frac{ \tilde{ \Omega }_\Lambda ( \bd{ k } ) }{ \left( \beta \omega \right)^2 } .
\end{equation}
The flow equations \eqref{eq:Sigmaflow3} and \eqref{eq:Uflow3}
then reduce to
\begin{widetext}
\begin{subequations}
\begin{align}
\partial_\Lambda \Sigma_\Lambda
= {} &
\frac{ 5 }{ 6 } U_\Lambda \frac{ T }{ N } \sum_{ \bd{q} } \dot{ G }_\Lambda ( \bd{q} )
+ \frac{ 2 }{ N T } \sum_{ \bd{q} } 
\frac{ \partial_\Lambda J_\Lambda ( \bd{q} ) }{ G_\Lambda ( \bd{q + Q} ) }
S_1 \left( \tilde{ \Omega }_\Lambda ( \bd{ q } ) , 
\tilde{ \Omega }_\Lambda ( \bd{ q }  + \bd{Q} ) \right) , 
\\
\partial_\Lambda U_\Lambda
= {} &
\frac{ T }{ N } \sum_{ \bd{q} } \dot{ G }_\Lambda ( \bd{q} ) 
\left[
\frac{ 7 }{ 10 } V_0 - \frac{ 11 }{ 3 } U_\Lambda^2 G_\Lambda ( \bd{q} )
\right]
-\frac{ 4 }{ N} \sum_{ \bd{q} } \left[ \partial_\Lambda J_\Lambda ( \bd{q} ) \right]
\left[
\frac{ b_3 }{ b_1^4 } S_2 \left( \tilde{ \Omega }_\Lambda ( \bd{ q } ) \right) 
\right]
\nonumber\\
&
- \frac{ 24}{ N T^3b_1^4} \sum_{ \bd{q} } 
\frac{ \partial_\Lambda J_\Lambda ( \bd{q} ) }{ 
G_\Lambda ( \bd{q} ) G_\Lambda^2 ( \bd{q} + \bd{Q} ) }
 S_4 \left( \tilde{ \Omega }_\Lambda ( \bd{ q } ) , 
\tilde{ \Omega }_\Lambda ( \bd{ q }  + \bd{Q} ) \right) ,
\end{align}
\end{subequations}
We have three distinct Matsubara sums appearing in this expression. Together with an auxiliary sum $S_3(x,y)$ to calculate $S_4(x,y)$ they are for $ x , y \ge 0$ given by
\begin{subequations}
\begin{align}
S_1 ( x , y )
= {} &
\sum_{ \omega \neq 0 } \frac{ \left( \beta \omega \right)^4 }{ 
\bigl[ \left( \beta \omega \right)^2 + x \bigr]^2
\bigl[ \left( \beta \omega \right)^2 + y \bigr] }
\nonumber\\
= {} &
\frac{ 1 }{ 8 \left( x - y \right)^2 }
\Bigl[
2 \sqrt{ x } \left( x - 3 y \right) \coth \left( \sqrt{ x } / 2 \right) 
%
- x \left( x - y \right) {\rm csch}^2 \left( \sqrt{ x } / 2 \right)
%
+ 4 y^{ 3 / 2 } \coth \left( \sqrt{ y } / 2 \right) 
\Bigr] ,
\\
S_2 ( x )
= {} &
\sum_{ \omega \neq 0 } \frac{ \left( \beta \omega \right)^2 }{ 
\bigl[ \left( \beta \omega \right)^2 + x \bigr]^2 }
= 
\frac{ \sqrt{x} - \sinh \left( \sqrt{x} \right)  }{ 
4 \sqrt{x} \left[ 1 - \cosh \left( \sqrt{x} \right) \right] } ,
\\
S_3 ( x , y )
= {} &
\sum_{ \omega \neq 0 } \frac{ \left( \beta \omega \right)^2 }{ 
\bigl[ \left( \beta \omega \right)^2 + x \bigr]^2
\bigl[ \left( \beta \omega \right)^2 + y \bigr] }
\nonumber\\
= {} &
\frac{ 1 }{ 8 \sqrt{ x } \left( x - y \right)^2 }
\Bigl[
2 \left( x + y \right) \coth \left( \sqrt{ x } / 2 \right) 
%
+ \sqrt{ x } \left( x - y \right) {\rm csch}^2 \left( \sqrt{ x } / 2 \right)
%
- 4 \sqrt{ x y } \coth \left( \sqrt{ y } / 2 \right)
\Bigr] ,
\\
S_4 ( x , y )
= {} &
\sum_{ \omega \neq 0 } \frac{ \left( \beta \omega \right)^6 }{ 
\bigl[ \left( \beta \omega \right)^2 + x \bigr]^3
\bigl[ \left( \beta \omega \right)^2 + y \bigr]^2 }
%
= 
S_3 ( x , y ) + \frac{ x }{ 2 } \partial_x S_3 ( x , y )
+ y \partial_y S_3 ( x , y ) 
%
+ \frac{ x y }{ 2 } \partial_x \partial_y S_3 ( x , y ) .
\end{align}
\end{subequations}
\end{widetext}

\begin{figure}
 \centering
\includegraphics[width=\linewidth]{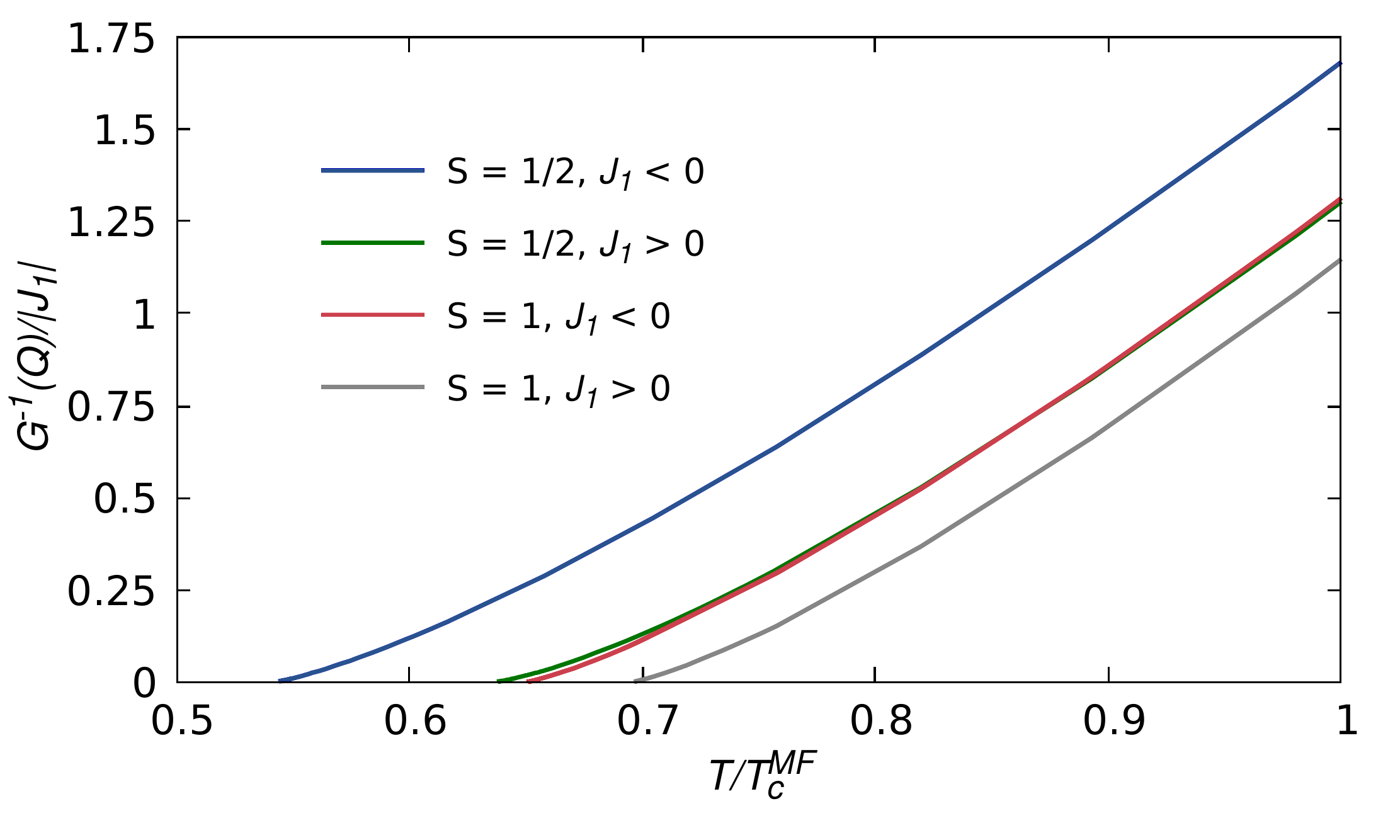}
\caption{Temperature dependence of the inverse spin susceptibility $G^{-1}(\bm{Q})$ 
including dynamic spin fluctuations, 
for quantum Heisenberg Models with only nearest neighbor interaction $J_1$ 
and $S = 1/2, 1$, 
using the interaction-switch deformation scheme \eqref{eq:switch}. 
}
\label{fig:dynamicsincluded}
\end{figure}

In Fig.~\ref{fig:dynamicsincluded}  we show our numerical results for $G^{-1}(\bm{Q})$ 
as a function of $T/T^{MF}_c$ for nearest-neighbor Heisenberg models with spin $S=1/2$ and $S=1$
using the interaction-switch deformation scheme.
We clearly observe a different effect of the finite-frequency diagrams 
with momentum transfer $\bm{Q} = \bm{\Gamma} =  (0,0,0)$ and $\bd{Q} =  \bm{R} = ( \pi , \pi , \pi)$:
While for the ferromagnet 
($J_1 <0$) quantum fluctuations enhance the spin self-energy and hence increase $T_c$,
in the case of an antiferromagnet
 ($J_1>0$) these fluctuations reduce the spin self-energy and thus 
lower $T_c$.
This should be contrasted with the static truncation of Sec.~\ref{sec:static},
which  could not distinguish between these two cases.

The transition temperatures of various quantum spin models are collected 
and compared to their benchmark values in Table \ref{table_3}. 
\begin{table}
\centering
\begin{tabular}{ c c c c c c c c c }
\hline
\hline
&&&&& \multicolumn{2}{c}{$T_c / T_c^{\rm MF}$} && \multicolumn{1}{c}{rel. error / $\%$} \\
\cline{6-7}
\cline{9-9}
$S$				&&	 $J_1$		&	$J_3 / J_1$	&&	switch	&	benchmark	&&	switch	\\
\hline
$1/2$ 		&&	$<0$			&	0						&&	0.545 	&	0.559			&&	2.5	\\
$1/2$ 		&&	$>0$			&	0						&&	0.640		&	0.629			&&	1.7	\\
$1$ 			&&	$<0$			&	0						&&	0.651		&	0.650			&&	0.2	\\
$1$				&&	$>0$			&	0						&&	0.697		&	0.684			&&	1.9	\\
$3/2$ 		&&	$<0$			&	0		&&	0.688	&	0.685		&&	0.4	\\
$3/2$ 		&&	$>0$			&	0		&&	0.715	&	0.702		&&	1.9\\
$1/2$ 		&&	$>0$			&	0.2	&&		0.752		&	0.722		&&	4.2		\\
$1/2$ 		&&	$>0$			&	0.4	&&	0.799	&	0.768		&&	4.0	\\
$1/2$ 		&&	$>0$			&	0.6	&&	0.823	&	0.794		&&	3.7		\\
$1/2$ 		&&	$>0$			&	0.8	&&	0.834			&	0.808		&&	3.2		\\
\hline
\hline
\end{tabular}
\caption{
Same as Tables \ref{table_1} and \ref{table_2},
but including dynamical (quantum) spin fluctuations.
Note that we do not list the classical magnet with $ S \to \infty $,
because quantum fluctuations vanish in this case.
Hence, it reduces to the static level-2 truncation discussed in Sec.~\ref{sec:level-2}.
}
\label{table_3}
\end{table}
We focus on the interaction-switch deformation scheme in this section
since it has proven more accurate in the static truncation (Sec.~\ref{sec:static})
and is numerically cheaper to implement when finite momentum transfer is involved.
We note that for quantum magnets which  order at finite $T_c$,
our spin FRG results have a similar or even higher accuracy than the pseudofermion FRG \cite{Iqbal16, Niggemann22}, 
with far less numerical overhead.
At the same time, 
the spin FRG flow equations depend only on correlation functions of the physical spins
and therefore enable us to devise physically motivated approximation schemes.

\section{Summary and outlook}

\label{sec:summary}

In this work we have used a particular implementation \cite{Tarasevych21} 
of the functional renormalization group approach to quantum spin systems \cite{Krieg19} 
to calculate the phase diagram and the critical temperatures of the $J_1 J_2 J_3$ quantum Heisenberg model on a cubic lattice. Recently this model has been used as a benchmark to test
different implementations of the pseudofermion FRG \cite{Ritter22}.
Within a rather simple static level-1 truncation of the hierarchy of the spin FRG flow equations,
we have obtained the critical temperature 
with a similar accuracy  as the numerically more expensive pseudofermion FRG 
for all values of the exchange couplings for which we have found controlled benchmark values in the literature.
Furthermore,
our spin FRG allows us to consider quantum spin systems at arbitrary spin quantum number $S \ge 1/2$
without any additional technical or numerical cost,
unlike the pseudofermion \cite{Baez2017} or pseudo-Majorana \cite{Niggemann2021} FRG implementations.

Away from the classical phase boundaries
where the ground states become degenerate,
we have developed more sophisticated truncations
that include the flow of the four-point vertex as well as dynamic (quantum) fluctuations
to obtain improved estimates for the critical temperature.
A comparison with available Monte-Carlo and high-temperature expansion results 
shows that our estimates for $T_c$
deviate at most by a few percent from the correct results. 
Moreover, our spin FRG approach allows us to  explicitly construct the
renormalization group fixed point which controls the critical behavior 
in the vicinity of the magnetic phase transition.

In the parameter regime where
classically the energies of two or more ordered states are (almost) degenerate 
we expect that the momentum dependence of the spin self-energy and of 
the effective four-spin interaction 
cannot be neglected. 
In this context, 
the vertex expansion with momentum- and frequency-dependent vertices should be contrasted with approximations based on the derivative expansion \cite{Berges02,Kopietz10} such as the local potential approximation.
While the derivative expansion  can non-perturbatively describe the 
field dependence of the average effective action, it does not 
provide easily accessible information about the momentum dependence of vertices. Moreover, a possible 
non-trivial frequency dependence of quantum vertices, 
which is expected to be crucial for the formation of a quantum
spin liquid in frustrated systems, 
is also not readily available within  
truncations based on the derivative expansion. 
We therefore believe that a vertex expansion 
which treats both the momentum- and frequency-dependence of the vertices
on equal footing and in an unbiased manner
is better suited for frustrated quantum spin systems
than the derivative expansion.

In order to go beyond the static level-1 truncation
and to reveal the  renormalization group
fixed points controlling the critical behavior of such strongly frustrated systems, 
we should divide the Brillouin zone into a large number of sectors 
and solve the resulting system of 
coupled differential equations resulting from the discretization 
of the FRG flow equations
for the spin self-energy and the effective interactions given in
Eqs.~\eqref{eq:Sigmaflow} and \eqref{eq:flow4a}.
In such a highly  frustrated parameter regime
we expect that at low temperatures ($T \ll |J_i|$)  the additional dynamic diagrams 
will generate positive contributions of order $| J_i |$  to the spin self-energy and the four-point vertex 
which stabilize a paramagnetic state.
In fact, in three dimensions and  for small $ T / | J_i |$
the magnitude of all terms containing only the bare three-point vertex 
may be estimated as such,
with all residual diagrams being of subleading order in $T/|J_i|$. 
However,  as we saw for the non-frustrated antiferromagnet, 
a positive net sign of these terms is not guaranteed at the relevant ordering vectors,
so that the precise mechanism which  stabilizes a paramagnetic state at low temperatures
remains somewhat intransparent in our approach.
This also suggests that a generic formula for the flowing dynamic spin-susceptibility $\tilde \Pi_\Lambda(K)$ 
like a high-frequency limit may not suffice 
and that a more prudent ansatz or a proper flow equation 
for the momentum and frequency dependence of $\tilde \Pi_\Lambda(K)$ is required.
The numerical solution of these equations is beyond the scope of this work.

Our approach can also be used to study 
frustrated spin systems in two dimensions.
In this case, we expect that the chiral dynamic spin 
fluctuations discussed in Sec.~\ref{sec:dyn}
will play an important role to destabilize magnetic order 
and possibly lead to a spin-liquid phase at zero temperature. 
In fact, the calculation of the spin dynamics
using our spin FRG approach is a challenging problem on its own 
because in this case the
proper implementation of conservation laws is essential. 
In the high temperature limit $( T \gg | J_i |)$ this problem has been solved in 
Ref.~[\onlinecite{Tarasevych21}] where the spin FRG has been used to derive
an integral equation for the dynamic spin susceptibility.  
The feedback of the spin dynamics onto the thermodynamics 
of frustrated magnets deserves further attention.

\section*{Acknowledgements}
We would like to thank Bj\"{o}rn Sbierski explaining some technical details of  the pseudo-Majorana FRG to us.
This work was financially supported by the 
Deutsche Forschungsgemeinschaft (DFG, German Research Foundation) 
through project KO 1442/10-1.

\appendix

\setcounter{equation}{0}

\renewcommand{\theequation}{A\arabic{equation}}

\renewcommand{\appendixname}{APPENDIX}

\renewcommand{\thesection}{\Alph{section}}

\section{Flow equations in static approximation}  

\label{app:static}

\renewcommand{\theequation}{A\arabic{equation}}

\begin{figure}
\includegraphics[width=\linewidth]{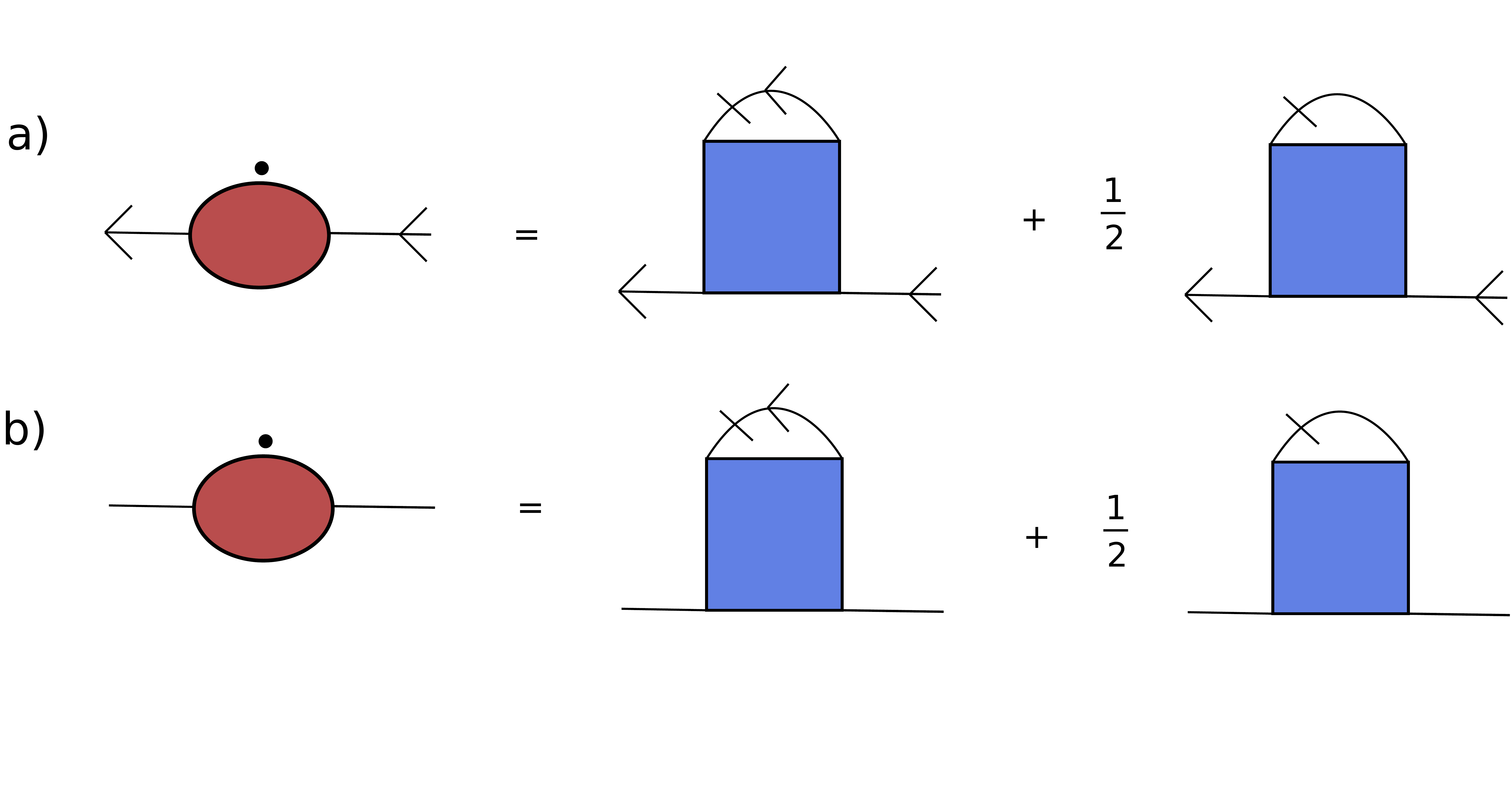}
\caption{%
Graphical representation of the flow equations 
\eqref{eq:flow2trans} (a) and \eqref{eq:flow2lon} (b) for the static spin self-energies. 
Arrows indicate transverse fluctuations $+, -$, 
a straight line represents a longitudinal degree of freedom $z$ and 
a slashed internal line is a static single-scale propagator $\dot G_\Lambda(\bm{q})$.
The dots above the left-hand sides denote the scale derivative.
}
\label{fig:diagrams2}
\end{figure}

In this Appendix we give the spin FRG flow equations for the
irreducible vertices defined via 
the hybrid functional $\Gamma_{\Lambda} [ \bd{m} , \bd{\eta} ]$
introduced in Sec.~\ref{sec:static} in static approximation, 
where the dynamical field $\bd{\eta}$ is set equal to zero.
For  a formal definition of the functional $\Gamma_{\Lambda} [ \bd{m} , \bd{\eta} ]$
see Ref.~[\onlinecite{Tarasevych21}].
Below we give the flow equations for a general spin $S$ Heisenberg model in a 
magnetic field $h$ with
Hamiltonian
\begin{equation}
{\cal H} = \frac{1}{2} \sum_{ij} J_{ij} \bd{S}_i \cdot \bd{S}_j -  h \sum_i S^z_i .
\end{equation}
To obtain the flow equations for the
$J_1 J_2 J_3$ Hamiltonian given in Eq.~\eqref{eq:hamiltonian} we should set $h=0$ and specify the
exchange couplings $J_{ij}$.
Since the external magnetic field breaks the spin-rotational invariance,
it is convenient to decompose  the magnetization field into a longitudinal component $m^z$
and into two spherical transverse components $ m^{\pm} = ( m^x   \pm i m^y ) / \sqrt{2}$.
In static approximation the vertex expansion of our deformed hybrid functional
is then \cite{Tarasevych21}
\begin{widetext}
\begin{align}
{\Gamma}_{\Lambda} [  \bd{m} , \bd{\eta} =0]   = {}&     
{\Gamma}_{\Lambda} [ 0 ,0 ]
+ \beta  \int_{\bd{k}} \left\{  
\left[ {J}_{ \bd{k} } +  {\Sigma}^{-+}_{\Lambda} ( \bd{k} ) \right]
m^-_{- \bd{k}} m^+_{\bd{k}} +
\frac{1}{2!}  \left[ {J}_{ \bd{k} } +  {\Sigma}^{zz}_{\Lambda} ( \bd{k} ) \right]
{m}^z_{- \bd{k} } {m}^z_{\bd{k}} 
\right\}
\nonumber\\
 &   
+ \beta  \int_{\bd{k}_1} \int_{\bd{k}_2} \int_{\bd{k}_3}  \int_{ \bd{k}_4} 
\delta ( \bd{k}_1 + \bd{k}_2 + \bd{k}_3 + \bd{k}_4) 
 \biggl\{  \frac{1}{ (2! )^2}
  \Gamma^{--++}_{\Lambda} ( \bd{k}_1, \bd{k}_2 ,  \bd{k}_3 ,  \bd{k}_4 )  
 m^-_{\bd{k}_1} m^-_{\bd{k}_2} m^+_{ \bd{k}_3 } m^+_{  \bd{k}_4} 
 \nonumber\\
&  
+ \frac{1}{ 2! } 
  \Gamma^{-+zz}_{\Lambda} ( \bd{k}_1, \bd{k}_2 ,  \bd{k}_3 ,  \bd{k}_4 )  
 m^-_{\bd{k}_1} {m}^+_{\bd{k}_2} m^z_{ \bd{k}_3 } m^z_{  \bd{k}_4} 
+  \frac{1}{4!} 
\Gamma^{zzzz}_{\Lambda} ( \bd{k}_1 , \bd{k}_2 , \bd{k}_3 , \bd{k}_4 ) 
 m^z_{\bd{k}_1 } m^z_{\bd{k}_2} m^z_{ \bd{k}_3 } 
 m^z_{\bd{k}_4} \biggr\}
 \nonumber\\
 & + \beta  \int_{\bd{k}_1} \ldots  \int_{ \bd{k}_6} 
\delta ( \bd{k}_1 + \ldots + \bd{k}_6) 
 \biggl\{  \frac{1}{ (3! )^2}
  \Gamma^{---+++}_{\Lambda} ( \bd{k}_1, \bd{k}_2 ,  \bd{k}_3 ,  \bd{k}_4 , \bd{k}_5 , \bd{k}_6)  
 m^-_{\bd{k}_1} m^-_{\bd{k}_2}   m^-_{\bd{k}_3} m^+_{ \bd{k}_4 } m^+_{  \bd{k}_5} m^+_{\bd{k}_6} 
 \nonumber\\
& 
+  \frac{1}{ (2 !)^3} \Gamma^{--++zz}_{\Lambda} ( \bd{k}_1, \bd{k}_2 ,  \bd{k}_3 ,  \bd{k}_4 , \bd{k}_5 , \bd{k}_6) 
m^-_{\bd{k}_1} m^-_{\bd{k}_2}   m^+_{\bd{k}_3} m^+_{ \bd{k}_4 } m^z_{  \bd{k}_5} m^z_{\bd{k}_6} 
\nonumber\\
&
 + \frac{1}{ 4 !} \Gamma^{-+zzzz}_{\Lambda} ( \bd{k}_1, \bd{k}_2 ,  \bd{k}_3 ,  \bd{k}_4 , \bd{k}_5 , \bd{k}_6) 
m^-_{\bd{k}_1} m^+_{\bd{k}_2}   m^z_{\bd{k}_3} m^z_{ \bd{k}_4 } m^z_{  \bd{k}_5} m^z_{\bd{k}_6}
\nonumber\\
&
 +  \frac{1}{ 6 !} \Gamma^{zzzzzz}_{\Lambda} ( \bd{k}_1, \bd{k}_2 ,  \bd{k}_3 ,  \bd{k}_4 , \bd{k}_5 , \bd{k}_6) 
m^z_{\bd{k}_1} m^z_{\bd{k}_2}   m^z_{\bd{k}_3} m^z_{ \bd{k}_4 } m^z_{  \bd{k}_5} m^z_{\bd{k}_6}
\biggr\}
+ \ldots \; , \hspace{7mm}
 \label{eq:Gammahcomplete}
\end{align}
\end{widetext}
where
 $\int_{\bd{k}} = \frac{1}{N} \sum_{\bd{k}}$.
Within this static truncation the
transverse spin self-energy $\Sigma_{\Lambda}^{-+} ( \bd{k} )$ satisfies
\begin{align}
\partial_{\Lambda} \Sigma_{\Lambda}^{-+} ( \bd{k} )  
= {} & 
\frac{T}{N} \sum_{\bd{q}} \Bigl[ 
\dot{G}^{+-}_{\Lambda} ( \bd{q} ) \Gamma^{--++}_{\Lambda} ( - \bd{k} , - \bd{q} , \bd{q} , \bd{k} )
\nonumber\\
& 
+ \frac{1}{2!}  \dot{G}^{zz}_{\Lambda} ( \bd{q} ) 
\Gamma^{-+zz}_{\Lambda} ( - \bd{k} ,  \bd{k} , - \bd{q} , \bd{q} ) \Bigr] ,
\label{eq:flow2trans}
\end{align}
while the flow of the longitudinal self-energy is
\begin{align}
\partial_{\Lambda} \Sigma_{\Lambda}^{zz} ( \bd{k} )  
= {} & 
\frac{T}{N} \sum_{\bd{q}} \Bigl[ 
\dot{G}^{+-}_{\Lambda} ( \bd{q} ) \Gamma^{-+zz}_{\Lambda} ( - \bd{q} ,  \bd{q} , - \bd{k} , \bd{k} )
\nonumber\\
& + \frac{1}{2!}  \dot{G}^{zz}_{\Lambda} ( \bd{q} ) 
\Gamma^{zzzz}_{\Lambda} ( - \bd{q} ,  \bd{q} , - \bd{k} , \bd{k} ) \Bigr] .
\label{eq:flow2lon}
\end{align}
Graphical representation of the flow equations \eqref{eq:flow2trans} and
\eqref{eq:flow2lon} are shown in Fig.~\ref{fig:diagrams2}.
For vanishing magnetic field and in the absence of a spontaneous magnetization,
the longitudinal spin self-energy agrees with the  the transverse one,
$\Sigma_{\Lambda}^{zz} ( \bd{k} ) = \Sigma^{-+}_{\Lambda} ( \bd{k} )
= \Sigma_{\Lambda} ( \bd{k} )$,
so that Eqs.~\eqref{eq:flow2trans} and \eqref{eq:flow2lon}
both reduce to the flow equation \eqref{eq:Sigmaflow} given in the main text. 
The relevant combinations of the four-point vertices is given by
\begin{align}
& 
\Gamma^{(4)}_{\Lambda} ( - \bd{q} ,  \bd{q} , - \bd{k} , \bd{k} ) 
\nonumber\\
= {} & 
\Gamma^{--++}_{\Lambda} ( - \bd{k} , - \bd{q} , \bd{q} , \bd{k} ) 
   +   \frac{1}{2!} \Gamma^{-+zz}_{\Lambda} ( - \bd{k} ,  \bd{k} , - \bd{q} , \bd{q} ) 
 \nonumber\\
= {} & 
\Gamma^{-+zz}_{\Lambda} ( - \bd{q} ,  \bd{q} , - \bd{k} , \bd{k} ) 
   +   \frac{1}{2!} \Gamma^{zzzz}_{\Lambda} ( - \bd{q} ,  \bd{q} , - \bd{k} , \bd{k} ) .
\label{eq:Gamma4rot}
\end{align}

\begin{figure*}
\includegraphics[width=\linewidth]{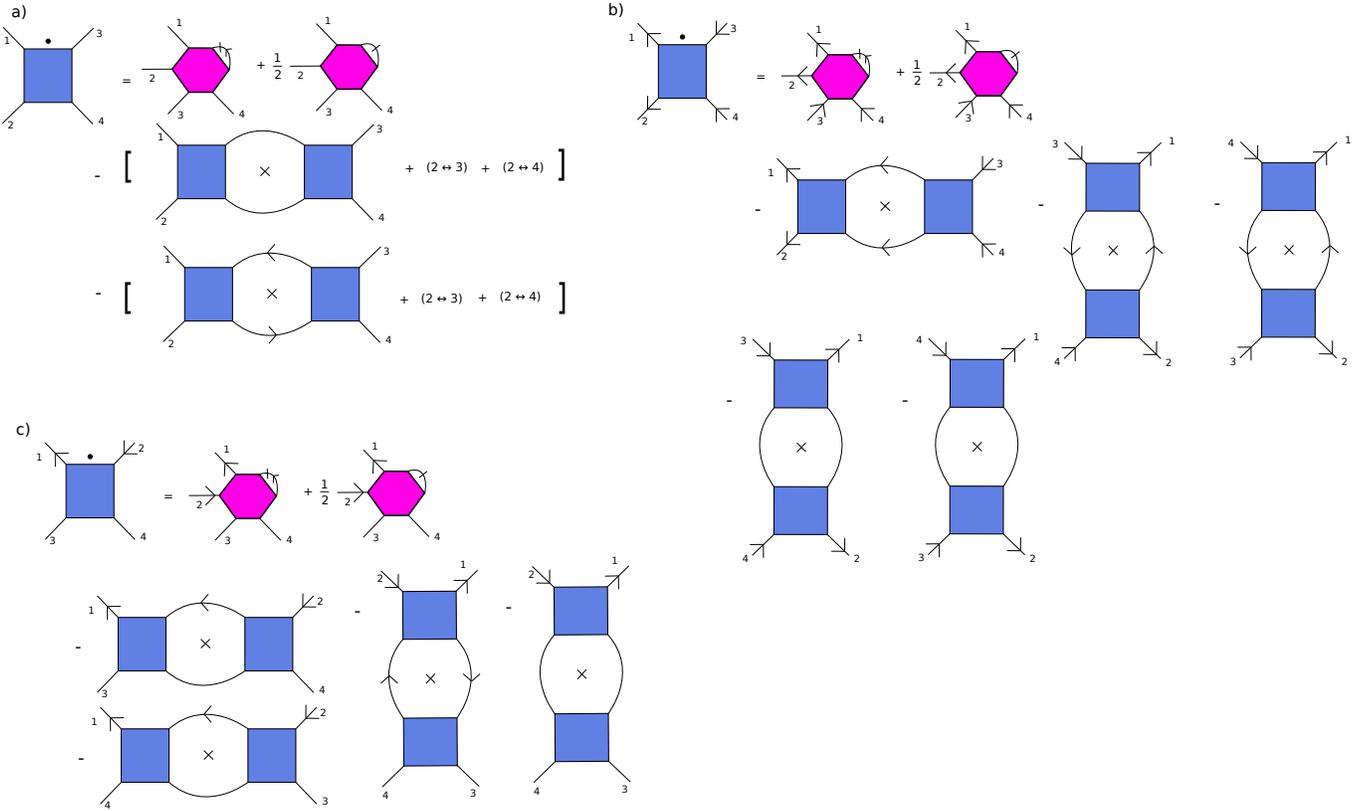}
\caption{%
Graphical representation of the flow equations  for the 
four-point vertices (a) $\Gamma^{zzzz}_\Lambda$, (b) $\Gamma^{++--}_\Lambda$, and
and (c) $\Gamma^{+-zz}_\Lambda$ in the static limit.
The notation is the same as in Figs.~\ref{fig:diagramsdyn} and \ref{fig:diagrams2}.
}
\label{fig:diagrams4}
\end{figure*}

The flow equations for the three types of static four-point vertices in an external magnetic field
are shown graphically  in Fig.~\ref{fig:diagrams4}.
Let us explicitly write down the flow equation for the
longitudinal four-point vertex,
\begin{widetext}
\begin{align}
& \partial_{\Lambda} \Gamma^{zzzz}_{\Lambda} ( \bd{k}_1 , \bd{k}_2 , \bd{k}_3 , \bd{k}_4 )  = 
 \frac{T}{N} \sum_{\bd{q}} \left[ 
 \dot{G}^{+-}_{\Lambda} ( \bd{q} ) 
 \Gamma^{-+zzzz}_{\Lambda} ( - \bd{q} , \bd{q} ,  \bd{k}_1 , \bd{k}_2 , \bd{k}_3 , \bd{k}_4 )
 + \frac{1}{2!} \dot{G}^{zz}_{\Lambda} ( \bd{q} ) 
 \Gamma^{zzzzzz}_{\Lambda} ( - \bd{q} , \bd{q} ,  \bd{k}_1 , \bd{k}_2 , \bd{k}_3 , \bd{k}_4 ) \right]
 \nonumber
 \\
 &  -   \frac{T}{N} \sum_{\bd{q}}  
 \Bigl\{ \dot{G}^{zz}_{\Lambda} ( \bd{q} ) G^{zz}_{\Lambda} ( \bd{q} + \bd{k}_1 + \bd{k}_2 )
 \Gamma_{\Lambda}^{zzzz} ( \bd{q} , - \bd{q} - \bd{k}_1 - \bd{k}_2 ,    \bd{k}_1 , \bd{k}_2  )
\Gamma_{\Lambda}^{zzzz} (   - \bd{q} , \bd{q} + \bd{k}_1 + \bd{k}_2 , \bd{k}_3 , \bd{k}_4 ) 
 \nonumber
 \\
 & \phantom{ -   \frac{T}{N} \sum_{\bd{q}} \Bigl\{  } 
+  ( \bd{k}_2 \leftrightarrow \bd{k}_3 ) + ( \bd{k}_2 \leftrightarrow \bd{k}_4 ) \Bigr\}
 \nonumber
 \\
&  -   \frac{T}{N} \sum_{\bd{q}} \Bigl\{ 
 \left[ {G}^{+-}_{\Lambda} ( \bd{q} ) G^{+-}_{\Lambda} ( \bd{q} + \bd{k}_1 + \bd{k}_2 )
 \right]^{\bullet}
 \Gamma_{\Lambda}^{-+zz} (  \bd{q} , - \bd{q} - \bd{k}_1 - \bd{k}_2 , \bd{k}_1 , \bd{k}_2   )
\Gamma_{\Lambda}^{-+zz} (   - \bd{q} , \bd{q} + \bd{k}_1 + \bd{k}_2 , \bd{k}_3 , \bd{k}_4 )
 \nonumber
 \\
 & \phantom{ -   \frac{T}{N} \sum_{\bd{q}} \Bigl\{  }
 + ( \bd{k}_2 \leftrightarrow \bd{k}_3 ) + ( \bd{k}_2 \leftrightarrow \bd{k}_4 ) \Bigr\} ,
 \label{eq:flow4a}
\end{align}
\end{widetext}
where we introduced the abbreviation
\begin{align}
\left[ 
G^{+-}_\Lambda ( \bd{q} ) G^{+-}_\Lambda ( \bd{q} + \bd{k} )
\right]^{\bullet}
= {} & 
\dot{G}^{+-}_\Lambda ( \bd{q} ) G^{+-}_\Lambda ( \bd{q} + \bd{k} ) 
\nonumber\\
&
+
G^{+-}_\Lambda ( \bd{q} ) \dot{G}^{+-}_\Lambda ( \bd{q} + \bd{k} ) .
\end{align}
For $h=0$,
spin-rotational symmetry implies that for vanishing external momenta, all  four-point vertices can be expressed 
in terms of a single scale-dependent coupling $U_{\Lambda}$ 
as follows,
 \begin{subequations}
\begin{align}
\Gamma^{zzzz}_{\Lambda} (0,0,0,0) & =  U_{\Lambda},
 \\
 \Gamma^{-+zz}_{\Lambda} (0,0,0,0) & =  \frac{1}{3} U_{\Lambda},
 \\
 \Gamma^{--++}_{\Lambda} (0,0,0,0) & =  \frac{2}{3} U_{\Lambda}.
 \end{align}
 \end{subequations}
Neglecting the momentum-dependence of all vertices and
keeping in mind that for vanishing external magnetic field and 
in the absence of spontaneous symmetry breaking 
$G^{+-}_{\Lambda} ( \bd{k} ) = G^{zz}_{\Lambda} ( \bd{k} ) = G_{\Lambda} ( \bd{k} )$,
the flow equation \eqref{eq:flow4a} for the longitudinal four-point vertex reduces to
 \begin{align}
 \partial_{\Lambda} U_{\Lambda} 
= & \frac{T}{N} \sum_{\bd{q}}
 \dot{G}_{\Lambda} ( \bd{q} ) 
 [ \Gamma_{\Lambda}^{-+zzzz} (0) + \frac{1}{2} 
 \Gamma_{\Lambda}^{zzzzzz}(0) ]
 \nonumber
 \\
- & \frac{11}{3} U_{\Lambda}^2 \frac{T}{N} \sum_{\bd{q}}
 \dot{G}_{\Lambda} ( \bd{q} ) {G}_{\Lambda} ( \bd{q} ) .
 \label{eq:flow4aa}
\end{align}
Similarly, the flow equation 
for the transverse four-point vertex shown in 
Fig.~\ref{fig:diagrams4} (b) reduces to
 \begin{align}
 \partial_{\Lambda} U_{\Lambda} 
= & 
\frac{T}{N} \sum_{\bd{q}}
 \dot{G}_{\Lambda} ( \bd{q} ) 
 \frac{3}{2} [ \Gamma_{\Lambda}^{---+++} (0) + \frac{1}{2} 
 \Gamma_{\Lambda}^{--++zz}(0) ]
 \nonumber
 \\
 & - \frac{11}{3} U_{\Lambda}^2 \frac{T}{N} \sum_{\bd{q}}
 \dot{G}_{\Lambda} ( \bd{q} ) {G}_{\Lambda} ( \bd{q} ) ,
 \label{eq:flow4bb}
\end{align}
while the flow equation  for the mixed four-point 
vertex in Fig.~\ref{fig:diagrams4} (c) reduces to
 \begin{align}
 \partial_{\Lambda} U_{\Lambda} 
& = \frac{T}{N} \sum_{\bd{q}}
 \dot{G}_{\Lambda} ( \bd{q} ) 
  3 [ \Gamma_{\Lambda}^{--++zz} (0) + \frac{1}{2} 
 \Gamma_{\Lambda}^{-+zzzz}(0) ]
 \nonumber
 \\
 & - \frac{11}{3} U_{\Lambda}^2 \frac{T}{N} \sum_{\bd{q}}
 \dot{G}_{\Lambda} ( \bd{q} ) {G}_{\Lambda} ( \bd{q} ) ,
 \label{eq:flow4cc}
\end{align}
Compatibility of Eqs.~\eqref{eq:flow4aa}, \eqref{eq:flow4bb} and 
\eqref{eq:flow4cc} implies that for $h=0$  the six-point vertices for vanishing momenta
satisfy
 \begin{align}
& 
 \Gamma_{\Lambda}^{-+zzzz} (0) + \frac{1}{2} 
 \Gamma_{\Lambda}^{zzzzzz}(0)
 \nonumber
 \\
 = {} &  \frac{3}{2} [ \Gamma_{\Lambda}^{---+++} (0) + \frac{1}{2} 
 \Gamma_{\Lambda}^{--++zz}(0) ]
 \nonumber
 \\
 = {} & 3 [ \Gamma_{\Lambda}^{--++zz} (0) + \frac{1}{2} 
 \Gamma_{\Lambda}^{-+zzzz}(0) ] .
 \label{eq:sixpointrel}
\end{align}
These are two independent relations between the four different types of six-point vertices. 
Thus, two of these vertices, for example the mixed vertices
$\Gamma_\Lambda^{--++zz}(0)$ and $\Gamma_\Lambda^{-+zzzz} (0)$, can be expressed
in terms of the purely longitudinal vertex
$\Gamma^{zzzzzz}_\Lambda (0)$ and  the transverse vertex $\Gamma^{---+++}_\Lambda (0)$.
We obtain
\begin{subequations}  \label{eq:sixrel2}
\begin{align}
 \Gamma_\Lambda^{--++zz} (0) & = \frac{2}{9} 
  \Gamma_\Lambda^{zzzzzz} (0) -  \frac{2}{9} \Gamma_\Lambda^{---+++}(0) ,
 \\
 \Gamma_\Lambda^{-+zzzz} (0) & = - \frac{1}{3} \Gamma_\Lambda^{zzzzzz} (0)
  + \frac{4}{3} \Gamma_\Lambda^{---+++}(0).
 \end{align}
\end{subequations}
The initial value of the purely longitudinal part of the six-point vertex can be expressed
in terms of the derivatives of the Brillouin function as 
follows \cite{Kriegthesis}:
 \begin{equation}
\Gamma^{zzzzzz}_{0} ( 0 )
 =   T \left(  - \frac{b_5}{b_1^6 }
  +   10  \frac{ b_3^2 }{ b_1^7 } \right) \equiv V_0.
 \label{eq:Gammazzzzzz}
\end{equation}
Using the generalized Wick theorem for spin operators derived in 
Ref.~[\onlinecite{Goll19}], 
we find that the transverse connected six-spin correlation function 
in a finite magnetic field $h$ is for vanishing frequencies given by
\begin{equation}
G_0^{+++---} (0) = \frac{6}{h^5} \left[ 
3 b ( y ) - 3 y b^{\prime} ( y ) + y^2 b^{\prime \prime} ( y ) 
 \right] ,
\end{equation}
where $y = \beta h$. 
For $h \rightarrow 0$ the term in the brackets vanishes as $y^5$, 
so that $G_0^{+++---} ( 0 )$ reduces to a finite constant in this limit,
 \begin{equation}
 G_0^{+++---} (0) = \frac{ 2}{5} b_5 \beta^5 ,
\end{equation}
where $b_5$ is given in Eq.~\eqref{eq:b5def}.
Using the tree expansion \cite{Kopietz10} that relates 
connected correlation functions to the irreducible vertices, 
we then obtain
 \begin{align} \label{eq:Gammammmppp}
\Gamma^{---+++}_{0} ( 0 ) 
&= T  \left( -  \frac{2}{5} \frac{b_5}{ b_1^6} + 4 \frac{ b_3^2}{b_1^7} \right)
\nonumber\\
& = \frac{2}{5} \Gamma_{0}^{zzzzzz}(0) = \frac{2}{5} V_0 .
\end{align}
Substituting our results~\eqref{eq:Gammazzzzzz} and \eqref{eq:Gammammmppp}
for the initial values of the longitudinal and transverse six-point vertices 
into the Eq.~\eqref{eq:sixrel2} we obtain for the mixed six-point vertices
at the initial scale,
\begin{align}
\Gamma^{--++zz}_{0} ( 0 ) & = \frac{2}{15} V_0 ,
 \\
\Gamma^{-+zzzz}_{0} ( 0 ) & = \frac{1}{5} V_0 .
\end{align}
The initial value of the combination of the six-point vertices 
that appears in the flow equation \eqref{eq:flow4aa} 
of the four-point vertex is therefore
\begin{align}
&
\Gamma_{0}^{-+zzzz} (0) + \frac{1}{2} 
 \Gamma_{0}^{zzzzzz}(0) 
\nonumber\\
= {} & \frac{7}{10} V_0 =
\frac{7 T}{ b_1^6} \left(  \frac{b_3^2}{b_1}   -  \frac{b_5}{10}    \right) > 0.
\end{align}

Note that relations between different types of $n$-point vertices 
for vanishing momenta can be generalized for finite momenta $\{\bd{k}_i\}$ using spin-rotational invariance. As a consequence we have only one  independent combination of spin components, for example $\Gamma^{++--}_\Lambda ( \bd{k}_1 ,
 \bd{k}_2 , \bd{k}_3, \bd{k}_4)$ for the 4-legged vertex. Using this type of relations will become important in the study of systems with frustrating interactions, where the momentum dependence of the self energy and the four-point vertices cannot be neglected.

\section{Fixed point}  

\label{app:fixed}

\renewcommand{\theequation}{B\arabic{equation}}

\begin{figure}
\centering
\includegraphics[width=\linewidth]{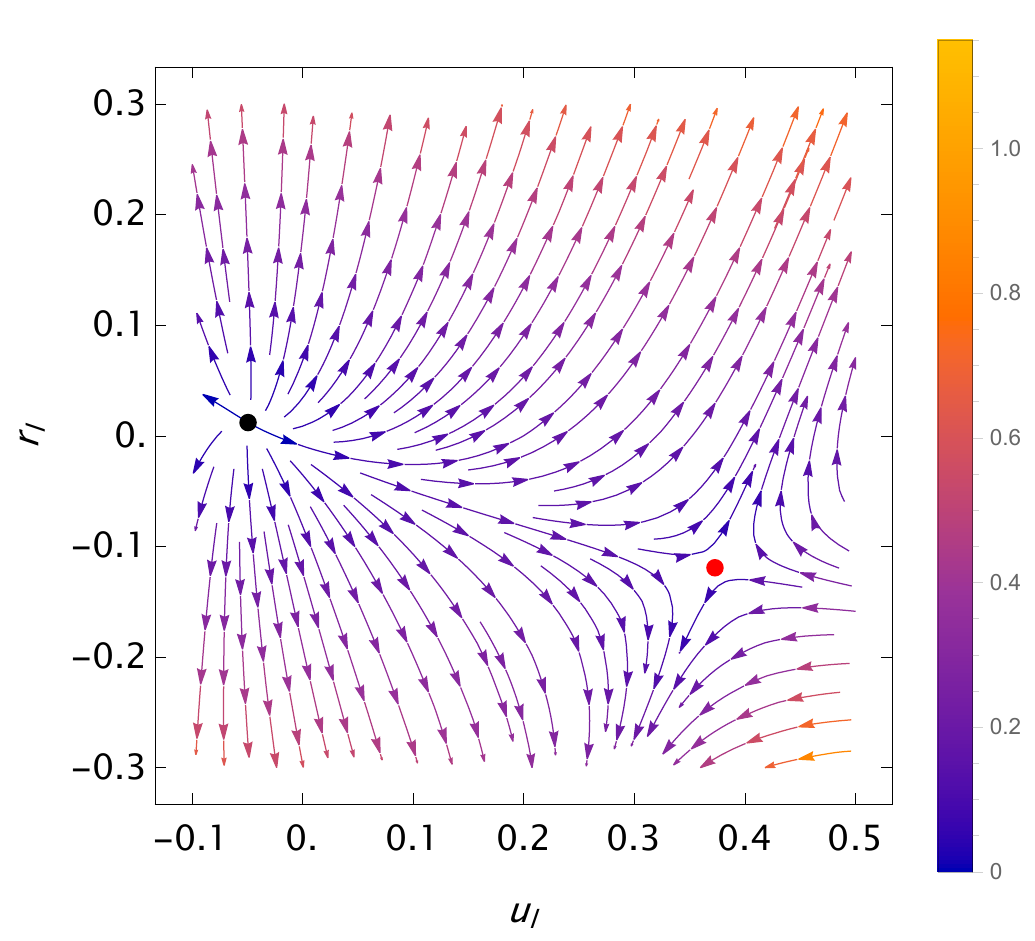}
\caption{%
Flow \eqref{eq:flow_rescaled} of the rescaled couplings for $l \to \infty$ in $D = 3$ dimensions,
for $S=1/2$ and temperature $T = 0.2\, T_{\rm MF}$.
The black and red points denote the Gaussian and Wilson-Fisher fixed points,
respectively.
}
\label{fig:flow}
\end{figure}

Any  finite-temperature continuous phase transition in the Heisenberg model 
can be associated with a critical fixed point of the renormalization group.
When the classical ground state is not degenerate, 
we furthermore expect that the critical fixed point 
can be identified with the usual Wilson-Fisher fixed point. 
In this Appendix,
we show how to recover this fixed point from the static spin FRG within the level-2 truncation.
To that end, 
we focus for simplicity on the nearest neighbor Heisenberg model,
such that $J_2 = 0 = J_3$. 
In this case $J_{ \bd{Q} } = J_{\rm min} = - J_{\rm max} = - 2D | J_1 |$ in $D$ dimensions,
and it is convenient to introduce dimensionless vertex functions as follows:
\begin{equation}
r_\Lambda = \frac{ \Sigma_\Lambda }{  2D  | J_1 | } - 1 , \;\;\;
u_\Lambda = \frac{ 5 }{ 6 } \frac{ U_\Lambda T }{ \left(  2D | J_1| \right)^2 } , \;\;\;
v_0 = \frac{ 7 }{ 6 } \frac{ V_0 T^2 }{ \left(  2D | J_1| \right)^3 } .
\end{equation}
With the Litim deformation scheme \eqref{eq:Litim},
the level-2 flow equations \eqref{eq:Sigmaflow_2} and \eqref{eq:Uflow2} then read
\begin{subequations} \label{eq:flow_Litim}
\begin{align} 
\partial_{\Lambda} r_{\Lambda} 
= {} &
- u_\Lambda I ( \Lambda )
\left[  
\frac{ 1 }{ \left( 1 + \Lambda + r_\Lambda \right)^2 } -
\frac{ 1 }{ \left(1 - \Lambda + r_\Lambda \right)^2 }
\right] ,
\\
\partial_{\Lambda} u_{\Lambda} 
= {} &
- \frac{ 1 }{ 2 } v_0  I ( \Lambda )
\left[  
\frac{ 1 }{ \left( 1 + \Lambda + r_\Lambda \right)^2 } -
\frac{ 1 }{ \left(1 - \Lambda + r_\Lambda \right)^2 }
\right]
\nonumber\\
&
+ \frac{ 22 }{ 5 } u_\Lambda^2  I ( \Lambda )
\left[  
\frac{ 1 }{ \left( 1 + \Lambda + r_\Lambda \right)^3 } -
\frac{ 1 }{ \left(1 - \Lambda + r_\Lambda \right)^3 }
\right] ,
\end{align}
\end{subequations}
where
\begin{align}
I ( \Lambda ) 
= {}  & 2D | J_1 | \int_\Lambda^1 d\epsilon\, 
\nu \left(  2D | J_1 | \epsilon \right)
\nonumber \\
=  {} & \frac{ 1 }{ N } \sum_{ \bd{q} } \Theta \left( J_{\bd{q} } -  2D | J_1 | \Lambda \right)
\end{align}
counts the number of states between the band edge and the deformation scale $\Lambda$.
To investigate the existence of a fixed point of the renormalization group flow,
we approximate $I ( \Lambda )$ by its asymptotic behavior for $\Lambda \to 1$,
\begin{equation}
I( \Lambda ) \sim I_D \left( 1 - \Lambda \right)^{D/2}, \; \; \; 1 - \Lambda \ll 1,
\label{eq:Ismall}
\end{equation}
where
\begin{equation}
I_D = \frac{ K_D }{ D } ( 2 D )^{D/2},
\label{eq:IDdef}
\end{equation}
and
\begin{equation}
K_D = \frac{1}{ 2^{D-1} \pi^{D/2} \Gamma (D/2) }
\end{equation}
is the surface area of the unit sphere divided by $(2 \pi )^D$.
Setting $\Lambda = 1 - e^{-2 l}$ and defining the rescaled couplings
\begin{equation}
r_l  =    e^{2l}  r_\Lambda  , \; \; \;
u_l = e^{ (4-D) l } u_\Lambda , \; \; \;
v_l = e^{ (6-2D) l } v_0 ,
\end{equation}
we find that the flow equations \eqref{eq:flow_Litim} are equivalent to the following system of equations:
\begin{subequations} \label{eq:flow_rescaled}
\begin{align}
\partial_l r_l 
= {} &
2 r_l - 2 u_l I_D \left[
\frac{ 1 }{ \left( 2 e^l - 1 + r_l \right)^2 } -
\frac{ 1 }{ \left( 1 + r_l \right)^2 }
\right] , 
\\
\partial_l u_l 
= {} &
( 4 - D ) u_l - v_l I_D \left[
\frac{ 1 }{ \left( 2 e^l - 1 + r_l \right)^2 } -
\frac{ 1 }{ \left( 1 + r_l \right)^2 }
\right] 
\nonumber\\
&
+ \frac{ 44 }{ 5 } u_l^2 I_D \left[
\frac{ 1 }{ \left( 2 e^l - 1 + r_l \right)^3 } -
\frac{ 1 }{ \left( 1 + r_l \right)^3 }
\right], 
\\
\partial_l v_l 
= {} &
( 6 - 2 D ) v_l .
\end{align}
\end{subequations}
The first terms in the square brackets on the right-hand sides of the above flow equations 
originate from the high-energy modes at the upper band edge
and vanish for $l \to \infty$.
These terms do not affect the fixed point.
The resulting flow at $l \to \infty$ is shown in Fig.~\ref{fig:flow},
where one clearly sees the Wilson-Fisher fixed point in addition to the Gaussian one. 
As expected, 
these equations are equivalent to the one-loop RG flow equations for the corresponding $\phi^4$-model, belonging to the $O(3)$-universality class \cite{Kopietz10}.

From the flow equations \eqref{eq:flow_rescaled},
we see that neglecting the flow of the four-point vertex results
in a runaway flow of the rescaled coupling $u_l$ for $D < 4$.
This is the reason why we do not obtain a true fixed point in the level-1 truncation of Sec.~\ref{sec:level-1}.
On the other hand,
for $D=3$ the rescaled six-point vertex $v_l$ is marginal and
does not flow within our approximation,  $v_l = v_0$,
thereby justifying the level-2 truncation used in Sec.~\ref{sec:level-2}. Note that the presence of a marginal coupling may introduce additional fixed points besides the Wilson-Fisher fixed point \cite{Yabunaka17}.

\section{Time-ordered correlation functions and irreducible vertices of a single spin}  

\label{app:single_spin}

\renewcommand{\theequation}{C\arabic{equation}}

In this Appendix we derive the irreducible mixed five-point and six-point vertices 
of a single spin given in Eqs.~\eqref{eq:5point} and \eqref{eq:6point}.
Therefore
we first calculate the imaginary-time ordered spin correlation functions, 
and then construct the corresponding irreducible vertices using the tree expansion.
In principle, all correlation functions of a single spin can be
obtained by means of the
generalized Wick theorem for spin operators \cite{Vaks68,Tarasevych21}.
In our case we only need correlation functions of the type
 \begin{equation}
G_0^{\overset{n}{\overbrace{z \ldots z}} +-} (0, \ldots ,0  ,  \omega , - \omega ) 
 \nonumber
 \end{equation}
involving $n$ operators $S^z$ at vanishing frequency and one operator pair $S^+$ and $S^-$ at 
finite frequencies.
These correlation functions can be obtained by taking $n$ derivatives of the magnetic field dependent
transverse two-point function
 \begin{equation}
 G_0^{+-} ( \omega , - \omega ) = \frac{b}{h - i \omega}
\end{equation}
with respect to the external magnetic field $h$ using the recursion relation \cite{Rueckriegel22}
\begin{align}
&  G_0^{\overset{n+1}{\overbrace{zz \ldots z}} +-} ( 0,0, \ldots ,0  ,  \omega , - \omega ) 
 \nonumber
 \\
= {} & 
\frac{\partial }{\partial h }  G_0^{\overset{n}{\overbrace{z \ldots z}} +-} ( 0, \ldots ,0  ,  \omega , - \omega ) .
 \end{align}
In particular, 
\begin{align}
  G_0^{z+-} ( 0 ,  \omega , - \omega ) 
	& =  \frac{ \partial }{\partial h} G_0^{+-} ( \omega , - \omega ) 
 \nonumber
 \\
 & =
- \frac{b}{ (h -  i \omega)^2 } 
+ \frac{ \beta b^{\prime}}{ h -  i \omega } ,
 \end{align}
and
\begin{align}
& G_0^{zz+-} ( 0 ,0,  \omega , - \omega )  
=   \frac{ \partial }{\partial h} G_0^{z+-} (0,  \omega , - \omega ) 
 \nonumber
 \\
& = \frac{2b}{ (h -  i \omega)^3 } -
 \frac{ 2\beta b^{\prime}}{ (h - i \omega)^2 }
 + \frac{\beta^2 b^{\prime \prime}}{ h - i \omega } .
 \end{align}
Taking one more $h$-derivative and then setting $h =0$ we obtain for vanishing magnetic field 
 \begin{align}
 G_0^{zzz+-} (0,0,0,  \omega , - \omega ) 
& = \frac{ 6\beta b_1}{ (-i \omega )^3}
 + \frac{ \beta^3 b_3}{-i \omega } 
 \nonumber
 \\
& = - \frac{\beta}{(  i \omega )^3}
 \left[ 6 b_1 -  b_3 ( \beta \omega )^2 \right] .
 \label{eq:Gzzzpm}
\end{align}
In this work we also need the 
mixed six-spin correlation function, 
which is  for $h \rightarrow 0$ given by
 \begin{align}
 G_0^{zzzz+-} (0,0,0,0,  \omega , - \omega ) 
& = 
 \frac{ 4}{i \omega } G_0^{zzz+-} (0,0,0,  \omega , - \omega )
 \nonumber
 \\
 & = 
 - \frac{4 \beta}{ \omega^4}
 \left[ 6 b_1 - b_3 ( \beta \omega )^2 \right] .
 \label{eq:G6mix}
\end{align}

\begin{figure}[tb]
 \includegraphics[width=\linewidth]{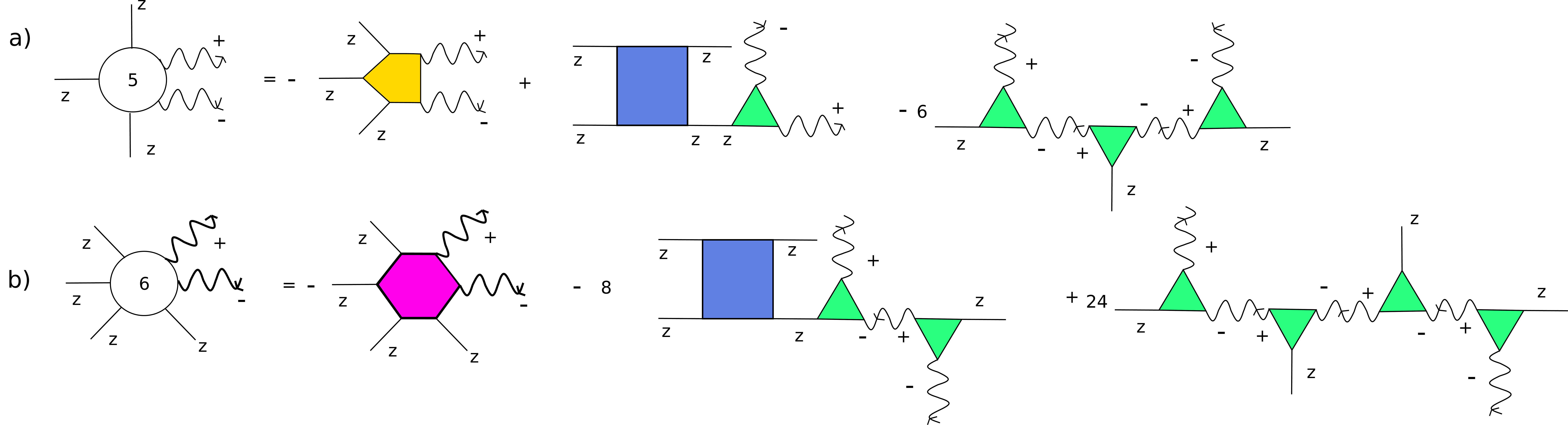}
  \caption{%
Tree expansion of the mixed five-spin and six-spin 
correlation functions $G_0^{zzz+-} ( 0,0,0, \omega , - \omega )$
and $G_0^{zzzz+-} ( 0, 0,0,0, \omega , - \omega )$ 
in terms of irreducible vertices.
}
\label{fig:treeexpansion}
\end{figure}

The corresponding irreducible vertices can be obtained from the tree expansion of
connected correlation functions in terms of irreducible vertices \cite{Kopietz10}.
For the relevant five-point function in zero magnetic field the tree expansion is
 \begin{align}
& G_0^{zzz+-} ( 0,0,0 , \omega , - \omega )  = 
 \nonumber
 \\
 &  - G_0^3 \Bigl[
 \Gamma_0^{ zzz \eta^- \eta^+} (0,0,0, - \omega , \omega ) 
 -   \Gamma_0^{zzzz} ( 0,0,0,0) G_0 
 \nonumber \\ 
 & 
\times \Gamma_0^{z \eta^- \eta^+} ( 0, - \omega , \omega ) 
 + 6 G^{-2}_0[\Gamma_0^{z \eta^- \eta^+} (0, - \omega , \omega )]^3
  \Bigr],
 \label{eq:tree5}
 \end{align}
where $G_0 = \beta b_1$ is the static spin susceptibility of an isolated spin.
A graphical representation of this relation is shown in  Fig.~\ref{fig:treeexpansion}.
Solving Eq.~\eqref{eq:tree5} for the mixed five-point vertex,
we obtain
 \begin{align}
 & \Gamma_0^{ zzz \eta^- \eta^+} (0,0,0, - \omega , \omega )  =
   \nonumber
 \\
 & 
  - G_0^{-3} G_0^{zzz+-} ( 0,0,0 , \omega , - \omega )
 +   \Gamma_0^{zzzz} ( 0,0,0,0) 
 \nonumber
 \\
 & 
 \times G_0 \Gamma_0^{z \eta^- \eta^+} ( 0, - \omega , \omega )
  - 6 G^{-2}_0[\Gamma_0^{z \eta^- \eta^+} (0, - \omega , \omega )]^3 ,
 \end{align}
Next, we substitute our result given in  Eq.~\eqref{eq:Gzzzpm} 
for  $G_0^{zzz+-} ( 0,0,0 , \omega , - \omega )$
and use
 \begin{equation}
\Gamma_0^{zzzz} ( 0,0,0,0) = - \frac{b_3 }{   \beta b_1^4}
 \end{equation}
 and
 \begin{equation}
 \Gamma_0^{z \eta^- \eta^+} ( 0, - \omega , \omega ) = \frac{1}{ i \omega } .
 \label{eq:3point2}
 \end{equation}
Then we find that all terms cancel so that
 \begin{equation}
 \Gamma_{0}^{zzz\eta^- \eta^+} ( 0,0,0, - \omega , \omega   )  =  
0,
 \end{equation}
which is Eq.~\eqref{eq:5point} of the main text.

Finally, consider the tree expansion of the initial value of the
mixed six-point vertex
\begin{align}
& G_0^{zzzz+-} (0, 0,0,0 , \omega , - \omega )  =
 \nonumber
 \\
 &  -  G_0^4
 \Bigl[
 \Gamma_0^{ zzzz \eta^- \eta^+} (0,0,0,0, - \omega , \omega ) 
 \nonumber
 \\
 & 
 - 8\Gamma^{zzzz}_0(0,0,0,0)[\Gamma_0^{z \eta^- \eta^+} (0, - \omega , \omega )]^2
 \nonumber \\ & 
 - 24 [\Gamma_0^{z \eta^- \eta^+} (0, - \omega , \omega )]^4(-G_0)^{-3}
 \Bigr],
 \nonumber
 \\
 &
 \label{eq:tree6}
 \end{align}
Solving for the mixed six-point vertex yields
\begin{align}
& \Gamma_0^{ zzzz \eta^- \eta^+} (0,0,0,0, - \omega , \omega ) = 
 \nonumber
 \\ &
 - G_0^{-4}G_0^{zzzz+-} (0, 0,0,0 , \omega , - \omega ) 
 + 8\Gamma^{zzzz}_0(0,0,0,0)
 \nonumber
 \\ &
 \times  [\Gamma_0^{z \eta^- \eta^+} (0, - \omega , \omega )]^2
 + 24 [\Gamma_0^{z \eta^- \eta^+} (0, - \omega , \omega )]^4(-G_0)^{-3}.
 \end{align}
Substituting our result \eqref{eq:G6mix} for
$G_0^{zzzz+-} (0, 0,0,0 , \omega , - \omega ) $ as well as
Eqs.~\eqref{eq:5point} and \eqref{eq:3point2} for the lower order vertices,
we obtain for the mixed six-point vertex
 \begin{equation}
\Gamma_{0}^{zzzz\eta^- \eta^+} ( 0,0,0,0, - \omega , \omega   ) = 
\frac{ 4  b_3  }{ \beta b_1^4 \omega^2},
 \label{eq:6pointapp}
 \end{equation}
which is Eq.~\eqref{eq:6point} of the main text.

\section{Integral equation for the dynamic susceptibility}  

\label{app:integral}

\renewcommand{\theequation}{D\arabic{equation}}

In the main text,
we have used the leading term \eqref{eq:Pi_high-T} of the high-temperature expansion 
to estimate the irreducible dynamic spin-susceptibility $\tilde{\Pi}_\Lambda ( K )$.
In this Appendix, 
we explore a more sophisticated ansatz,
given by 
\begin{equation} 
\tilde{ \Pi } ( \bd{k} , i\omega )
= G ( \bd{k} ) \frac{ \Delta ( \bd{k} , i\omega ) }{ | \omega | } ,
 \label{eq:ansatzDelta}
\end{equation}
where $\Delta ( \bd{k} , i\omega )$ is the dissipation energy introduced in
Ref.~[\onlinecite{Tarasevych21}]. 
In the high-temperature limit the dissipation energy
satisfies the approximate integral equation \cite{Tarasevych21}
\begin{equation} \label{eq:Delta_integral}
\Delta ( \bd{k} , i\omega ) 
= \frac{ 1 }{ N } \sum_{ \bd{q} } \frac{ V ( \bd{k} , \bd{q} ) 
}{ \Delta ( \bd{k} , i\omega ) + | \omega | } .
\end{equation}
Here,
the kernel is given by
\begin{align}
V ( \bd{k} , \bd{q} ) 
= {} &
\frac{ b_1 }{ 4 } 
\left[  
\left( J_{ \bd{q} } - J_{ \bd{q} + \bd{k} } \right)^2 +
\left( J_{ \bd{q} } - J_{ \bd{q} - \bd{k} } \right)^2
\right]
\nonumber\\
&
+ 2 \Sigma_2 ( \bd{q} ) - \Sigma_2 ( \bd{q} + \bd{k} ) - \Sigma_2 ( \bd{q} - \bd{k} ) ,
\end{align}
where
\begin{equation}
\Sigma_2 ( \bd{k} ) 
= \frac{ 1 }{ N } \sum_{ \bd{q} } 
\left(
\frac{ J_{ \bd{q} } J_{ \bd{q} + \bd{k} } }{ 12 } -
\frac{ 5 }{ 6 } \frac{ b_3 }{ b_1 } J_{ \bd{q} }^2
\right)
\end{equation}
is the momentum-dependent part of the static spin self-energy
to leading order in a high-temperature expansion.
The flowing $\tilde{ \Pi }_\Lambda ( \bd{k} , i\omega )$
is then obtained by replacing the exchange coupling by
its deformed counterpart,
$ J_{ \bd{k} } \to J_\Lambda ( \bd{k} ) $.
Compared to the high-temperature approximation \eqref{eq:Pi_high-T} 
used in the main text,
this ansatz implies a non-trivial frequency dependence of 
$\tilde{ \Pi }_\Lambda ( \bd{k} , i\omega )$
that deviates from the $ \omega^{ - 2 } $-behavior for sufficiently small $\omega$.
For instance, in three dimensions one finds $\tilde{ \Pi }_\Lambda ( \bd{k} , i\omega ) \propto \bd{k}^2 / | \omega |$
implying spin diffusion \cite{Tarasevych21}. 
Another advantage of the ansatz (\ref{eq:ansatzDelta})  is
that it remains finite in the limit $ T \to 0 $
for any constant frequency.
Thus,
it may prove useful for the investigation of possible spin-liquid phases.
However,
one should keep in mind that similar 
to the outright high-temperature approximation \eqref{eq:Pi_high-T},
a high-temperature limit also underlies the validity 
of the integral equation \eqref{eq:Delta_integral} \cite{Tarasevych21}.
Therefore, 
we likewise expect it to break down at the end of the flow for $\Lambda \to 1$
in the regime $ T \lesssim | J_{ \bd Q } | $. 
A downside of this ansatz is furthermore
that the Matsubara sums in the flow equations \eqref{eq:Sigmaflow3} and \eqref{eq:Uflow3}
can no longer be performed analytically.
For the explicit numerical evaluation,
we therefore used a cutoff $| \omega_{\rm max} | = 50 \pi T$
beyond which all terms are neglected.
Convergence of the sums was confirmed by comparing with results computed with twice that cutoff, 
$| \omega_{\rm max} | = 100 \pi T$.

\begin{figure}
 \centering
\includegraphics[width=\linewidth]{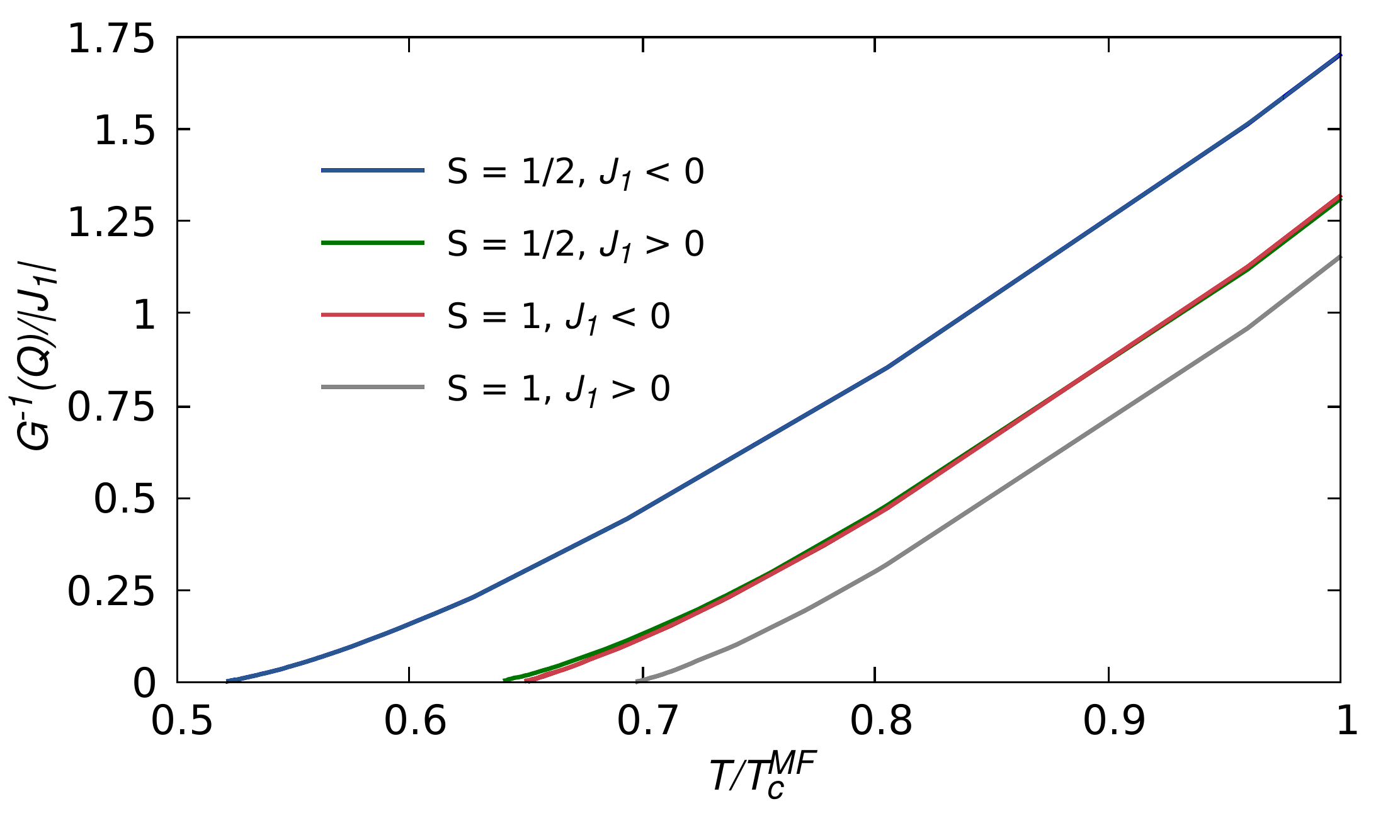}
\caption{Temperature dependence of the inverse spin susceptibility $G^{-1}(\bm{Q})$ 
using the integral equation \eqref{eq:Delta_integral}
to estimate the dynamic spin- susceptibility $\tilde{\Pi}_\Lambda ( K )$.
The plot is for Heisenberg models with spin $S=1/2$ and $S=1$ 
and nearest neighbor interaction $J_1$ 
using the interaction-switch deformation scheme \eqref{eq:switch}. 
}
\label{fig:integral}
\end{figure}

The inverse spin susceptibility $G^{-1}(\bm{Q})$ obtained 
with the integral equation \eqref{eq:Delta_integral}
is displayed in Fig.~\ref{fig:integral}
using the interaction-switch deformation scheme to integrate the flow equations.
For simplicity we focus on nearest neighbor Heisenberg magnets 
with  spin $S = 1/2$ and $S=1$.
The respective critical temperatures and their relative deviations from the benchmark values
are shown in Table \ref{table_4}.
\begin{table}
\centering
\begin{tabular}{ c c c c c c c c c }
\hline
\hline
&&&&& \multicolumn{2}{c}{$T_c / T_c^{\rm MF}$} && \multicolumn{1}{c}{rel. error / $\%$} \\
\cline{6-7}
\cline{9-9}
$S$				&&	 $J_1$		&	$J_3 / J_1$	&&	switch	&	benchmark	&&	switch	\\
\hline
$1/2$ 		&&	$<0$			&	0						&&	0.521 	&	0.559			&&	6.8 		\\
$1/2$ 		&&	$>0$			&	0						&&	0.641		&	0.629			&&	1.9			\\
$1$ 			&&	$<0$			&	0						&&	0.649		&	0.650			&& 0.2			\\
$1$				&&	$>0$			&	0						&&	0.698 	&	0.684			&&  2.0			\\
\hline
\hline
\end{tabular}
\caption{
Same as Table \ref{table_3}, but now the self-consistent ansatz for $\tilde \Pi_\Lambda (K)$
based on the integral equation \eqref{eq:Delta_integral} is employed. Values are listed only for the cases shown in Fig.~\ref{fig:integral}.
}
\label{table_4}
\end{table}
Compared to the result of Sec.~\ref{sec:dyn}
that employed the high-temperature approximation \eqref{eq:Pi_high-T}
for the dynamic spin-susceptibility,
the results are quantitatively comparable.
The only major deviation is the $S=1/2$ nearest-neighbor ferromagnet,
where the self-consistent ansatz (\ref{eq:ansatzDelta}) actually performs worse.


\begin{thebibliography}{99}
%
\bibitem{Diep20}
H.~T.~Diep, Editor, 
{\it Frustrated Spin Systems}, 
(World Scientific, Singapore, 3rd Edition, 2020).
%
\bibitem{White92}
S.~R.~White, 
{\it Density Matrix Formulation for Quantum Renormalization Groups},
\href{https://doi.org/10.1103/PhysRevLett.69.2863}
{Phys.~Rev.~Lett.~{\bf 69}, 2863 (1992)}.
%
\bibitem{Schollwoeck05}
U.~Schollw\"{o}ck, 
{\it The density-matrix renormalization group}, 
\href{https://doi.org/10.1103/RevModPhys.77.259}
{Rev.~Mod.~Phys.~{\bf 77}, 259 (2005)}.
%
\bibitem{Wetterich93}
C.~Wetterich, 
{\it Exact evolution equation for the effective potential}, 
\href{https://doi.org/10.1016/0370-2693(93)90726-X}
{Phys.~Lett.~B {\bf 301}, 90 (1993)}.
%
\bibitem{Berges02}
J.~Berges, N.~Tetradis, and C.~Wetterich,
{\it Non-perturbative renormalization flow in quantum field theory and statistical physics},
\href{https://doi.org/10.1016/S0370-1573(01)00098-9}
{Phys.~Rep.~{\bf 363}, 223 (2002)}.
%
\bibitem{Pawlowski07}
J.~M.~Pawlowski, 
{\it Aspects of the functional renormalisation group},
\href{https://doi.org/10.1016/j.aop.2007.01.007}
{Ann.~Phys.~{\bf 322}, 2831 (2007)}.
%
\bibitem{Kopietz10}
P.~Kopietz, L.~Bartosch, and F.~Sch\"{u}tz, 
{\it Introduction to the Functional Renormalization Group}, 
(Springer, Berlin, 2010).
%
\bibitem{Metzner12}
W.~Metzner, M.~Salmhofer, C.~Honerkamp, V.~Meden, and K.~Sch\"{o}nhammer,
{\it Functional renormalization group approach to correlated fermion systems},
\href{https://doi.org/10.1103/RevModPhys.84.299}
{Rev.~Mod.~Phys.~{\bf 84}, 299 (2012)}.
%
\bibitem{Dupuis21}
N.~Dupuis, L.~Canet, A.~Eichhorn, W.~Metzner, J.~M.~Pawlowski, M.~Tissier, and N.~Wschebor, 
{\it{The nonperturbative functional renormalization group and its applications}}, 
\href{https://doi.org/10.1016/j.physrep.2021.01.001}
{Phys.~Rep.~{\bf 910}, 1 (2021)}.
%
\bibitem{Hille20}
C. Hille, F. B. Kugler, C. J. Eckhardt, Y.-Y. He, A. Kaush, C. Honerkamp, A. Toschi, and
S. Andergassen, 
{\it{Quantitative functional renormalization group description of the two-dimensional 
Hubbard model}},
\href{https://doi/10.1103/PhysRevResearch.2.033372}
{Phys. Rev. Research {\bf{2}}, 033372 (2020).}
%
\bibitem{Abrikosov65}
A.~A.~Abrikosov, 
{\it Electron scattering on magnetic impurities in metals and anomalous resistivity effects},
\href{https://doi.org/10.1103/PhysicsPhysiqueFizika.2.5}
{Physics {\bf 2}, 5 (1965)}.
%
%
%
%
\bibitem{Reuther10}
J.~Reuther and P.~W\"{o}lfle, 
{\it $J_1$-$J_2$ frustrated two-dimensional Heisenberg model: 
Random phase approximation and functional renormalization group},
\href{https://doi.org/10.1103/PhysRevB.81.144410}
{Phys.~Rev.~B {\bf 81}, 144410 (2010)}.
%
\bibitem{Reuther11}
J.~Reuther and R.~Thomale,
{\it Functional renormalization group for the anisotropic triangular antiferromagnet },
\href{https://doi.org/10.1103/PhysRevB.83.024402}
{Phys.~Rev.~B {\bf 83}, 024402 (2011)}.
%
\bibitem{Reuther11a}
J.~Reuther, R.~Thomale, and S.~Trebst,
{\it Finite-temperature phase diagram of the Heisenberg-Kitaev model},
\href{https://doi.org/10.1103/PhysRevB.84.100406}
{Phys.~Rev.~B {\bf 84}, 100406(R) (2011)}.
%
\bibitem{Buessen16}
F.~L.~Buessen and S.~Trebst,
{\it Competing magnetic orders and spin liquids in two- and three-dimensional kagome systems: Pseudofermion functional renormalization group perspective},
\href{https://doi.org/10.1103/PhysRevB.94.235138}
{Phys.~Rev.~B {\bf 94}, 235138 (2016)}.
%
\bibitem{Iqbal16}
Y.~Iqbal, R.~Thomale, F.~P.~Toldin, S.~Rachel, and J.~Reuther,
{\it Functional renormalization group approach for three-dimensional quantum magnetism},
\href{https://doi.org/10.1103/PhysRevB.94.140408}
{Phys.~Rev.~B {\bf 94}, 140408(R) (2016)}.
%
\bibitem{Baez2017}
M.~L.~Baez and J.~Reuther,
{\it Numerical treatment of spin systems with unrestricted spin length $S$: 
A functional renormalization group study},
\href{https://doi.org/10.1103/PhysRevB.96.045144}
{Phys.~Rev.~B {\bf 96}, 045144 (2017)}.
%
\bibitem{Rueck18}
M.~R\"{u}ck and J.~Reuther, 
{\it Effects of two-loop contributions in the pseudofermion functional renormalization group method
for quantum spin systems},
\href{https://doi.org/10.1103/PhysRevB.97.144404}
{Phys.~Rev.~B {\bf 97}, 144404 (2018)}.
%
\bibitem{Thoenniss20}
J. Thoenniss, M. K. Ritter, F. B. Kugler, J. von Delft, and M. Punk,
{\it{Multiloop pseudofermion functional renormalization for quantum spin systems:
Application to the spin-$1/2$ kagome Heisenberg model}}, 
\href{https://doi.org/10.48550/arXiv.2011.01268}
{arXiv:2011.01268v1 [cond-mat.str-el] 2 Nov 2020}.
%
\bibitem{Kiese22}
D.~Kiese, T.~M\"{u}ller, Y.~Iqbal, R.~Thomale, and S.~Trebst,
{\it{Multiloop functional renormalization group approach to quantum spin systems}},
\href{https://doi.org/10.1103/PhysRevResearch.4.023185}
{Phys.~Rev.~Research {\bf 4}, 023185 (2022)}.
%
\bibitem{Ritter22}
M.~K.~Ritter, D.~Kiese, T.~M\"{u}ller, F.~B.~Kugler, R.~Thomale, S.~Trebst, and J.~von Delft,
{\it Benchmark Calculations of Multiloop Pseudofermion fRG}, 
\href{https://doi.org/10.1140/epjb/s10051-022-00349-2}
{Eur.~Phys.~J.~B {\bf 95}, 102 (2022)}.
%
%
%
%
%
\bibitem{Martin1959}
J.~L.~Martin, 
{\it Generalized classical dynamics, and the ‘classical analogue’ of a Fermioscillator},
\href{https://doi.org/10.1098/rspa.1959.0126}
{Proc.~R.~Soc.~London A {\bf 251}, 536 (1959)}.
%
\bibitem{Tsvelik1992}
A.~M.~Tsvelik, 
{\it New fermionic description of quantum spin liquid state},
\href{https://doi.org/10.1103/PhysRevLett.69.2142}
{Phys.~Rev.~Lett.~{\bf 69}, 2142 (1992)}.
%
\bibitem{Niggemann2021}
N.~Niggemann, B.~Sbierski, and J.~Reuther,
{\it Frustrated quantum spins at finite temperature: Pseudo-Majorana functional renormalization group approach},
\href{https://doi.org/10.1103/PhysRevB.103.104431}
{Phys.~Rev.~B {\bf 103}, 104431 (2021)}.
%
\bibitem{Niggemann22}
N.~Niggemann, J.~Reuther and B.~Sbierski,
{\it Quantitative functional renormalization for three-dimensional quantum Heisenberg models}, 
\href{https://doi.org/10.21468/SciPostPhys.12.5.156}
{SciPost Phys.~{\bf 12}, 156 (2022)}.
%
%
%
\bibitem{Krieg19}
J.~Krieg and P.~Kopietz, 
{\it Exact renormalization group for quantum spin systems}, 
\href{https://doi.org/10.1103/PhysRevB.99.060403}
{Phys.~Rev.~B {\bf 99}, 060403(R) (2019)}.
%
\bibitem{Tarasevych18}
D.~Tarasevych, J.~Krieg, and P.~Kopietz,
{\it A rich man's derivation of scaling laws for the Kondo model},
\href{https://doi.org/10.1103/PhysRevB.98.235133}
{Phys.~Rev.~B {\bf 98}, 235133 (2018)}.
%
\bibitem{Goll19}
R.~Goll, D.~Tarasevych, J.~Krieg, and P.~Kopietz,
{\it Spin functional renormalization group for quantum Heisenberg ferromagnets: 
Magnetization and magnon damping in two dimensions}, 
\href{https://doi.org/10.1103/PhysRevB.100.174424}
{Phys.~Rev.~B {\bf 100}, 174424 (2019)}.
%
\bibitem{Goll20}
R.~Goll, A.~R\"{u}ckriegel, and P.~Kopietz,
{\it Zero-magnon sound in quantum Heisenberg ferromagnets},
\href{https://doi.org/10.1103/PhysRevB.102.224437}
{Phys.~Rev.~B {\bf 102}, 224437 (2020)}.
%
\bibitem{Tarasevych21}
D.~Tarasevych and P.~Kopietz, 
Dissipative spin dynamics in hot quantum paramagnets,
\href{https://doi.org/10.1103/PhysRevB.104.024423}
{Phys.~Rev.~B {\bf 104}, 024423 (2021)}.
%
\bibitem{Tarasevych22}
D.~Tarasevych and P.~Kopietz, 
{\it Critical spin dynamics of Heisenberg ferromagnets revisited}, 
\href{https://doi.org/10.1103/PhysRevB.105.024403}
{Phys.~Rev.~B {\bf 105}, 024403 (2022)}.
%
\bibitem{Rueckriegel22}
A.~R\"{u}ckriegel, J.~Arnold, R.~Goll, and P. Kopietz,
{\it Spin functional renormalization group for dimerized quantum spin systems},
\href{https://doi.org/10.1103/PhysRevB.105.224406}
{Phys.~Rev.~B {\bf 105}, 224406 (2022)}.
%
%
%
%
\bibitem{Rancon11a}
A.~Ran\c{c}on and N.~Dupuis, 
{\it Nonperturbative renormalization group approach to the Bose-Hubbard model},
\href{https://doi.org/10.1103/PhysRevB.83.172501}
{Phys.~Rev.~B {\bf 83}, 172501 (2011)}.
%
\bibitem{Rancon11b}
A.~Ran\c{c}on and N.~Dupuis,  
{\it Nonperturbative renormalization group approach to strongly correlated lattice bosons},
\href{https://doi.org/10.1103/PhysRevB.84.174513}
{Phys.~Rev.~B {\bf 84}, 174513 (2011)}.
%
\bibitem{Rancon12b}
A.~Ran\c{c}on and N.~Dupuis,  
{\it Thermodynamics of a Bose gas near the superfluid-Mott-insulator transition},
\href{https://doi.org/10.1103/PhysRevA.86.043624}
{Phys.~Rev.~A {\bf 86}, 043624 (2012)}.
%
\bibitem{Rancon14}
A.~Ran\c{c}on, 
{\it Nonperturbative renormalization group approach to quantum XY spin models}, 
\href{https://doi.org/10.1103/PhysRevB.89.214418}
{Phys.~Rev.~B {\bf 89}, 214418 (2014)}.
%
%
%
%
\bibitem{Litim01}
D.~F.~Litim, 
{\it Optimized renormalization group flows},
\href{https://doi.org/10.1103/PhysRevD.64.105007}
{Phys.~Rev.~D {\bf 64}, 105007 (2001)}.
%
%
%
%
\bibitem{Vaks68}
V.~G.~Vaks, A.~I.~Larkin, S.~A.~Pikin,
{\it Thermodynamics of an Ideal Ferromagnetic Substance},
Zh.~Eksp.~Teor.~Fiz.~{\bf 53}, 281 (1967) 
[
\href{http://jetp.ras.ru/cgi-bin/e/index/e/26/1/p188?a=list}
{Sov.~Phys.~JETP {\bf 26}, 188 (1968)}
].
%
\bibitem{Vaks68b}
V.~G.~Vaks, A.~I.~Larkin, and S.~A.~Pikin, 
{\it Spin waves and correlation functions in a ferromagnetic}, 
Zh.~Eksp.~Teor.~Fiz.~{\bf 53}, 1089 (1967) 
[
\href{http://jetp.ras.ru/cgi-bin/e/index/e/26/3/p647?a=list}
{Sov.~Phys.~JETP {\bf 26}, 647 (1968)}
].
%
\bibitem{Izyumov88}
Yu. A. Izyumov and Yu. N. Skryabin,
{\it{Statistical Mechanics of Magnetically Ordered Systems}}, (Springer, Berlin, 1988).
%
%
%
%
\bibitem{Machado10}
T.~Machado and N.~Dupuis, 
{\it From local to critical fluctuations in lattice models: A nonperturbative renormalization-group approach},
\href{https://doi.org/10.1103/PhysRevE.82.041128}
{Phys.~Rev.~E {\bf 82}, 041128 (2010)}.
%
\bibitem{Meden02}
V.~Meden, W.~Metzner, U.~Schollw\"{o}ck, and K.~Sch\"{o}nhammer,
{\it Scaling behavior of impurities in mesoscopic Luttinger liquids},
\href{https://doi.org/10.1103/PhysRevB.65.045318}
{Phys.~Rev.~B {\bf 65}, 045318 (2002)}.
%
%
\bibitem{footnote_classical}
The classical limit is obtained by rescaling
$ J_{ \bd{ k } } S^2 \to J_{ \bd{ k } } $,
such that the Heisenberg Hamiltonian \eqref{eq:hamiltonian}
remains finite in the $S \to \infty$ limit.
The vertex functions used throughout this work then need to be rescaled according to
$ \Sigma_\Lambda S^2 \to \Sigma_\Lambda $,
$ U_\Lambda S^4 \to U_\Lambda $, and
$ V_\Lambda S^6 \to V_\Lambda $,
with initial conditions
$ \Sigma_0 = T / 3 $,
$ U_0 = 54 T / 5 $, and
$ V_0 = 7128 T / 35 $.
%
%
\bibitem{Oitmaa04}
J.~Oitmaa and W.~Zheng,
{\it Curie and N\'{e}el temperatures of quantum magnets},
\href{https://doi.org/10.1088/0953-8984/16/47/016}
{J.~Phys.~Condens.~Matter {\bf 16}, 8653 (2004)}.
%
\bibitem{Troyer04}
M.~Troyer, F.~Alet, and S.~Wessel,
{\it Histogram Methods for Quantum Systems: 
from reweighting to Wang-Landau Sampling},
\href{https://www.scielo.br/j/bjp/a/NbwwJnKFN7KB9kbrDrvsXtr/?lang=en}
{Braz.~J.~Phys.~{\bf 34}, 377 (2004)}.
%
\bibitem{Sandvik98}
A.~W.~Sandvik, 
{\it Critical Temperature and the Transition from Quantum to Classical 
Order Parameter Fluctuations in the Three-Dimensional Heisenberg Antiferromagnet},
\href{https://doi.org/10.1103/PhysRevLett.80.5196}
{Phys.~Rev.~Lett.~{\bf 80}, 5196 (1998)}.
%
\bibitem{Mao03}
W. Mao, P. Coleman, C. Hooley, and D. Langreth,
{\it{Spin Dynamics from Majorana Fermions}},
\href{https://doi/10.1103/PhysRevLett.91.207203}
{Phys. Rev. Lett. {\bf{91}}, 207203 (2003)}.
%
\bibitem{Shnirman03}
A. Shnirman and Y. Makhlin,
{\it{Spin-Spin Correlators in the Majorana Representation}},
\href{https://doi/10.1103/PhysRevLett.91.207204}
{Phys. Rev. Lett. {\bf{91}}, 207204 (2003).}
%
\bibitem{Kriegthesis}
J.~Krieg, 
{\it Functional renormalization group approach to classical and quantum spin systems},
(PhD-Thesis, Goethe-Universit\"{a}t Frankfurt, 2019).
%
\bibitem{Yabunaka17}
S. Yabunaka and B. Delamotte,
{\it Surprises in \textit{O(N)} Models: Nonperturbative Fixed Points, Large N Limits, and Multicriticality}, \href{https://doi.org/10.1103/PhysRevLett.119.191602} {Phys. Rev. Lett. {\bf{119}}, 191602 (2017).}


%
%
%
%
%
%
%
%
%
%
%
%
%
%
%
%
%
%
%
%
%
%
%
\end{thebibliography}
\end{document}